\documentclass[12pt,english,longbibliography,nofootinbib,superscriptaddress,12pt,sort&compress,showkeys]{article}
\usepackage{ae,aecompl}
\usepackage[T1]{fontenc}
\usepackage[latin9]{inputenc}
\usepackage[active]{srcltx}
\usepackage{float}
\usepackage{multirow}
\usepackage{amsmath}
\usepackage{graphicx}
\usepackage[numbers]{natbib}

\makeatletter

\providecommand{\tabularnewline}{\\}

\newcommand{\lyxaddress}[1]{
	\par {\raggedright #1
	\vspace{1.4em}
	\noindent\par}
}

\usepackage{aecompl}\usepackage{epstopdf}

\usepackage[raggedrightboxes]{ragged2e}

\usepackage{babel}

\makeatother

\usepackage{babel}
\begin{document}
\title{First-principles prediction of point defect energies and concentrations
in the tantalum and hafnium carbides}
\author{I. Khatri, R. K. Koju, and Y. Mishin}
\maketitle

\lyxaddress{Department of Physics and Astronomy, MSN 3F3, George Mason University,
Fairfax, Virginia 22030, USA}
\begin{abstract}
\noindent First-principles calculations are combined with a statistical-mechanical
model to predict the equilibrium point-defect concentrations in the
refractory carbides TaC and HfC as a function of temperature and chemical
composition. Several different types of point defects (vacancies,
interstitials, antisite atoms) and their clusters are treated in a
unified manner. The defect concentrations either strictly follow or
can be closely approximated by Arrhenius functions with parameters
predicted by the model. The model is general and applicable to other
carbides, nitrides, borides, or similar chemical compounds. Implications
of this work for understanding the diffusion mechanisms in TaC and
HfC are discussed. 
\end{abstract}

\section{Introduction\label{sec:Introduction}}

The refractory carbides TaC and HfC belong to the class of ultra-high
temperature ceramics (UHTCs), which includes several other transition
metal carbides, nitrides, and borides. UHTCs are characterized by
a high melting temperature, large elastic moduli, large hardness,
good thermal resistance, and relatively low chemical reactivity. The
TaC and HfC carbides have the highest melting temperatures $T_{\mathrm{m}}$
among all UHTCs. In fact, HfC has the highest melting temperature
(about 3942$^{\circ}$C) of all materials known today, with TaC's
$T_{\mathrm{m}}$ being only slightly lower (about 3900$^{\circ}$C)
\citep{Cedillos-Barraza:2016tp}. 

Both carbides have the B1-ordered (NaCl prototype) crystal structure
with significant deviations from the 50-50 stoichiometry towards carbon-deficient
compositions. This off-stoichiometry is accommodated by constitutional
(structural) vacancies on the carbon sublattice. The chemical bonding
in TaC and HfC combines three contributions: metallic bonding due
to the presence of the metallic atoms, covalent metal-carbon bonds,
and some degree of ionic bonding caused by partial metal-carbon charge
transfer \citep{Vines:2005vu}. The covalent bonds are the strongest,
making both carbides mechanically strong, hard and brittle, leading
to the high melting temperatures. 

The TaC and HfC carbides are notoriously difficult to sinter due to
the extremely small diffusion coefficients. The rate of pore healing
during the sintering is kinetically controlled by diffusive mass transport.
Thus, the knowledge of diffusion coefficients in TaC and HfC is essential
for optimizing the synthesis and processing routes. Experimental information
about the diffusivity of either carbon or the metallic atoms in these
carbides is scarce and indirect. Diffusion measurements are highly
challenging as they must be conducted at temperatures exceeding 2000-2800$^{\circ}$C
to generate reliable concentration curves from which to extract the
diffusion coefficients. Carbon diffusion in TaC was estimated by back-calculation
from the growth rate of an oxide layer \citep{Resnick1966,Brizes:1968aa}
or carbide layer \citep{Rafaja:1998aa}. Interdiffusion coefficients
in TaC and HfC were extracted from evaporation data \citep{Wallace:1996aa}.
No experimental data is available for metal self-diffusion in TaC
or HfC.

Under the circumstances, calculations offer the only realistic option
for obtaining the diffusivities in these carbides (and perhaps all
other carbides of the UHTC family). Point-defect concentrations constitute
an essential ingredient for the diffusion calculations. Density-functional
theory (DFT) calculations have been performed for vacancies and vacancy
clusters in TaC \citep{Tang:2020wb,Jubair:2019uj,Sun:2020uf,De-Leon:2015ud,Yu:2014uh,Yan:2016wj,Rimsza:2019vp,Vines:2005vu,Yu:2015us,Salehin:2021ti}
and HfC \citep{Yu:2015us,Salehin:2021ti}. Based on these and other
calculations \citep{Tang:2020wb,Jubair:2019uj,Sun:2020uf,De-Leon:2015ud,Yu:2014uh,Yu:2015us,Razumovskiy:2015wx},
it is assumed that both carbon and metallic atoms in TaC and HfC diffuse
by vacancy-atom exchanges on the respective sublattices. Among other
findings, it was predicted that the binding energy between the carbon
and metal vacancies in HfC is much stronger than in TaC, leading to
the formation of vacancy clusters in which a Hf vacancy is surrounded
by several carbon vacancies \citep{Salehin:2021ti}. However, thermodynamically
consistent calculations of the point-defect (and point-defect cluster)
formation energies and equilibrium concentrations have been challenging.
Such calculations must consider that point defects can only appear
and disappear by pairs or clusters that preserve the chemical composition.
Furthermore, different defect clusters must be in equilibrium with
each other with respect to composition-conserving dissociation-recombination
reactions. The only consistent treatment known to us was for vacancies
and antisite defects in the Ti, Zr and Hf carbides and nitrides \citep{Razumovskiy:2015wx}. 

This paper aims to predict point-defect concentrations in TaC and
HfC as a function of temperature and deviation from stoichiometry.
To this end, we develop a methodology that combines DFT calculations
with statistical mechanics accounting for the point-defect energies
and entropies, including both configurational and orientational entropy
contributions. This allows us to predict the equilibrium concentrations
of single point defects and point-defect clusters of any complexity.
The methodology is general enough to apply to other B1-ordered carbides,
nitrides, or borides in the future.

\section{Methodology}

The point defect energies were obtained by DFT calculations using
the Vienna Ab initio Simulation package (VASP) \citep{Kresse1996,Kresse96a}.
The calculations were carried out using six different cubic or orthorhombic
supercells containing 96, 144, 216, 288, 384, and 512 atoms, obtained
by replicating the conventional 8-atom unit cell of the B1 structure.
Using large supercell sizes with up to 512 atoms ensured size convergence,
which was especially important for large defect clusters creating
strong elastic strain fields. The convergence plots are presented
in the Supplementary Information file accompanying this article. The
projected augmented wave \citep{Blochl94} method was utilized with
exchange-correlation interactions treated in the generalized gradient
approximation in the Perdew-Burke-Ernzerhof formalism \citep{perdew92:gga_apps,PerdewBE96}.
The semi-core $p$ electrons were included as valence states for both
Ta and Hf. Following convergence test, the energy cutoffs of 780 eV
and 860 eV were applied to HfC and TaC, respectively. The k-point
grid convergence tests were conducted for all supercells; see the
Supplementary Information file. A Gaussian smearing of width 0.05
eV was chosen to perform electron minimization with the convergence
criterion of \textbf{$10^{-8}$} eV/atom. For ionic relaxations, we
used a conjugate-gradient algorithm with the 0.001 eV/\AA\ force criterion,
followed by a quasi-Newton algorithm to further improve accuracy.
Before introducing defects, each supercell was subjected to volumetric
relaxation while preserving its original shape. The point defects
were created by adding, removing, or changing the species of an atom
or a group of atoms, and the structures were relaxed with respect
to local atomic displacements.

Prior to the point-defect calculations, perfect lattice properties
of both carbides were computed to demonstrate the reliability of the
DFT methodology. Table \ref{table:lattice} shows that the lattice
parameter $a$ and elastic constants $c_{ij}$ obtained by our calculations
compare well with the available experimental \citep{brown1966elastic,weber1973lattice,Zhang:2017tx}
and theoretical \citep{he2008crystal,Yu:2014uh,Yu:2015us,guo2021electronic}
values. The elastic constants were computed by the energy-strain method
using second order polynomial fits for strains below 6\%.

The obtained energies of the perfect and defected supercells served
as input to the statistical-mechanical model for calculating the equilibrium
point-defect concentrations as explained below.

\section{Point-defect energies \label{sec:Point-defect-energies}}

The B1 structure of the TaC and HfC carbides consists of two penetrating
face-centered cubic (FCC) sublattices occupied by metallic and carbon
atoms. In the perfectly stoichiometric carbides at zero temperature,
the two sublattices are filled with the respective atoms without vacancies.
At finite temperatures, a stoichiometric carbide develops thermal
disorder in the form of vacancies on both sublattices, antisite defects,
interstitial atoms, and clusters of these defects. Deviations from
the perfect stoichiometry are accommodated by additional point defects
called constitutional (or structural). Depending on the chemical composition
and temperature, the disorder is dominated by either thermal or constitutional
defects. At low temperatures, the defects are primarily constitutional
and can be different on either side of the stoichiometric composition.
They can also be different between TaC and HfC.

To analyze the point defects in both carbides in a unified manner,
we will consider a generic carbide AB, in which the element A is either
Ta or Hf and the element B is carbon. The respective FCC sublattices
are denoted $\alpha$ and $\beta$. This generalized notation will
allow us to apply the present analysis to other binary carbides or
any ordered compound with the AB stoichiometry in the future.

There can be six types of elementary (single) point defects: 
\begin{itemize}
\item[] $V_{\alpha}$ = vacancy on sublattice $\alpha$ 
\item[] $V_{\beta}$ = vacancy on sublattice $\beta$ 
\item[] $A_{\beta}$ = antisite atom A on sublattice $\beta$ 
\item[] $B_{\alpha}$ = antisite atom B on sublattice $\alpha$ 
\item[] $I_{A}$ = interstitial metallic atom 
\item[] $I_{B}$ = interstitial carbon atom 
\end{itemize}
An antisite defect is obtained by replacing a metallic atom with carbon
on the metallic sublattice ($B_{\alpha}$) or a carbon atom with a
metallic atom on the carbon sublattice ($A_{\beta}$). Interstitial
atoms are inserted in a tetrahedral position.

The elementary defects can form dynamic clusters. The simplest cluster
is a pair of elementary defects separated by a nearest-neighbor distance
$r_{0}$. Examples include divacancies $V_{\alpha}V_{\beta}$, antisite
pairs $A_{\beta}B_{\alpha}$, and vacancy-antisite pairs such as $V_{\alpha}B_{\alpha}$
and $V_{\beta}A_{\beta}$. In the divacancy and the antisite pairs,
the elementary defects are nearest neighbors on different sublattices
($r_{0}=a/2$), whereas in the vacancy-antisite pairs $V_{\alpha}B_{\alpha}$
and $V_{\beta}A_{\beta}$, they are nearest neighbors on the same
sublattice ($r_{0}=a/\sqrt{2}$). Another defect pair, called a Frenkel
pair, is composed of a vacancy and interstitial atom. This defect
is obtained by moving an atom from its sublattice to an interstitial
position. Defect clusters can be composed of three or more elementary
defects and can have several geometric configurations.

DFT calculations produce a set of ``raw'' energies \citep{Mishin_97a,Mishin_97b,Mishin00b,Lozovoi03b}
of point defects, both elementary and clustered. The ``raw'' energy
of a point defect is \emph{defined} as the energy difference between
a relaxed supercell containing the defect and a perfect supercell
containing the same number of sites. Specifically, the ``raw'' energy
$\varepsilon_{d}$ of a defect $d$ is calculated by the formula 
\begin{equation}
\varepsilon_{d}=E_{d}(N)-E_{0}(N),\label{eq:raw-energy}
\end{equation}
where $E_{d}(N)$ and $E_{0}(N)$ are the total energies of an $N$-site
supercell with and without the defect, respectively. Since the two
supercells can have different chemical compositions, the ``raw''
energy generally depends on the reference atomic energy used in the
DFT calculations. Exceptions include the antisite pairs $A_{\beta}B_{\alpha}$,
Frenkel defects such as $V_{\alpha}I_{A}$ and $V_{\beta}I_{B}$,
and other defects obtained by displacing atoms from their perfect
lattice positions without changing the chemical composition. In all
other cases, the ``raw'' energy of the defect is not a physically
meaningful quantity by itself. However, it can be shown \citep{Mishin_97a,Mishin_97b,Mishin00b,Lozovoi03b}
that a complete set of ``raw'' energies, together with the cohesive
energy $\varepsilon_{0}$ of the perfect crystal (potential energy
per atom relative to ideal gas) uniquely defines the equilibrium point-defect
concentrations. Such concentrations are calculated using the statistical-mechanical
model discussed in section \ref{sec:Point-defect-concentrations},
which uses the ``raw'' energies as input.

\begin{figure}
\begin{centering}
\includegraphics[width=0.7\textwidth]{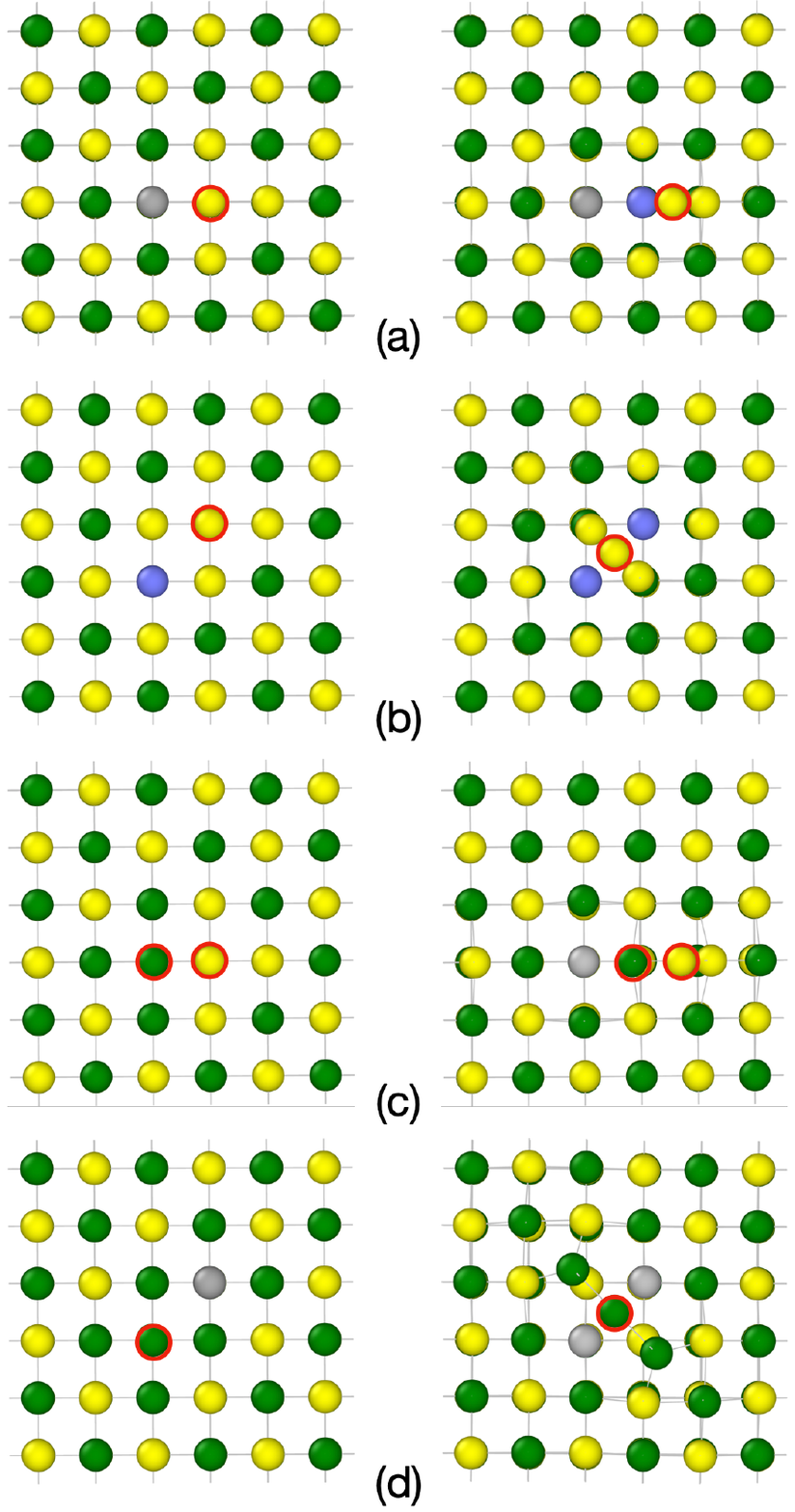} 
\par\end{centering}
\caption{Initial (left column) and relaxed (right column) structures of selected
point-defect pairs undergoing structural reconstructions. (a) $V_{\beta}B_{\alpha}$
pair in HfC (b) $V_{\alpha}B_{\alpha}$ pair in TaC and HfC. (c) $A_{\beta}B_{\alpha}$
antisite pair in TaC and HfC. (d) $V_{\beta}A_{\beta}$ pair in TaC
. The structures are viewed along a $\left\langle 100\right\rangle $
direction. The metal atoms, carbon atoms, metal vacancies, and carbon
vacancies are shown in green, yellow, purple, and light gray, respectively.
Some atoms are encircled in red for tracking.\label{fig:relaxed_structures}}
\end{figure}

Table \ref{table:raw} summarizes the ``raw'' energies of the elementary
defects and several defect clusters in both carbides. The ``raw''
energy of each defect was determined by linear extrapolation to zero
of the values obtained in different supercells when plotted against
the reciprocal of the number of sites, as proposed in Ref.~\citep{Mishin2001b}.
The plots are shown in the Supplementary Information file. Some of
the geometrically possible defect clusters are mechanically unstable
and are not included in Table \ref{table:raw}. For example, the vacancy-antisite
pair $V_{\alpha}A_{\beta}$ is unstable in both carbides: during the
relaxation, the antisite atom $A_{\beta}$ fills the vacancy and the
pair transforms into a carbon vacancy $V_{\beta}$. The vacancy-antisite
pair $V_{\beta}B_{\alpha}$ in TaC is also unstable and relaxes into
a metallic vacancy $V_{\alpha}$. However, in HfC, the same pair survives
relaxation and transforms into a linear structure consisting of a
divacancy $V_{\alpha}V_{\alpha}$ and a carbon interstitial dumbbell
$I_{B}$ aligned parallel to the {[}110{]} direction (Fig.~\ref{fig:relaxed_structures}(a)).
As another example, the metallic Frenkel pair $V_{\alpha}I_{A}$ comprising
a vacancy and a tetrahedral interstitial is unstable in both carbides:
during the relaxation, the interstitial atom fills the vacancy, recovering
the perfect crystal. The carbon Frenkel pair $V_{\beta}I_{B}$ is
likewise unstable in TaC but remains stable in HfC.

In some of the defect pairs, the relaxation is accompanied by small
atomic displacements preserving the initial elementary defects. For
example, the relaxed divacancies $V_{\alpha}V_{\beta}$ in both carbides
are composed of two distinct vacancies. In other cases, the defect
pair undergoes a significant reconstruction. For example, during the
relaxation of the $V_{\alpha}B_{\alpha}$ pair, the antisite carbon
atom $B_{\alpha}$ shifts into an interstitial position, leaving a
metallic vacancy behind.\footnote{This interstitial position is at the midpoint of the nearest-neighbor
C-C bond in the B1 structure and cannot be classified as either tetrahedral
or octahedral. It is unique to this reconstructed structure. } As a result, the initial defect pair relaxes into a linear structure
comprising a metallic divacancy $V_{\alpha}V_{\alpha}$ and a carbon
interstitial $I_{B}$ in between (Fig.~\ref{fig:relaxed_structures}(b)).
The antisite pair $A_{\beta}B_{\alpha}$ also reconstructs upon relaxation:
the antisite carbon atom $B_{\alpha}$ relaxes toward a nearby carbon
site while the antisite metallic atom $A_{\beta}$ relaxes toward
a metallic site, creating a linear structure consisting of a metallic
vacancy and a carbon interstitial dumbbell with the {[}110{]} orientation
(Fig.~\ref{fig:relaxed_structures}(c)). As yet another example,
the $V_{\beta}A_{\beta}$ pair in TaC relaxes into a carbon divacancy
$V_{\beta}V_{\beta}$ and a chain of three metallic atoms in a criss-cross
configuration (Fig.~\ref{fig:relaxed_structures}(d)).

In addition to the divacancy $V_{\alpha}V_{\beta}$, Table \ref{table:raw}
includes the ``raw'' energies of vacancy clusters $V_{\alpha}V_{\beta}^{n}$,
$n=2,...,6$, obtained by adding up to six carbon vacancies as nearest
neighbors of the metallic vacancy. The structures with $n=2$, $3$
and $4$ can have several symmetrically non-equivalent configurations
with different energies. The $V_{\alpha}V_{\beta}^{2}$ cluster can
have a triangular (T) or linear (L) configuration described in Ref.~\citep{Salehin:2021ti}.
In the $V_{\alpha}V_{\beta}^{3}$ cluster, the carbon vacancies can
form an in-plane (IP) or off-plane (OP) configuration \citep{Salehin:2021ti}.
In the $V_{\alpha}V_{\beta}^{4}$ cluster, the two occupied nearest-neighbor
carbon sites of $V_{\alpha}$ can be in either L or T configuration.
Accordingly, the four carbon vacancies can form an IP or an OP configuration
\citep{Salehin:2021ti}. By contrast, all configurations of the $V_{\alpha}V_{\beta}^{5}$
and $V_{\alpha}V_{\beta}^{6}$ clusters are symmetrically equivalent.
In the latter case, all nearest-neighbor sites of $V_{\alpha}$ are
vacant.

The binding energies of the elementary defects into the clusters were
also calculated. The binding energy is defined as the ``raw'' energy
difference between the cluster and a system of isolated elementary
defects forming the cluster. This energy can be calculated using a
supercell containing the cluster and a set of supercells containing
the elementary defects. The respective supercell energies must be
appropriately scaled to ensure the conservation of the total number
of sites \citep{Razumovskiy:2015wx}. In contrast to the ``raw''
energies, the binding energy is a well-defined physical quantity independent
of reference energies. A negative binding energy indicates that the
elementary defects attract each other when forming the cluster.

The binding energies obtained by the DFT calculations are reported
in Table \ref{table:bindingenergy}. Note that the divacancy and the
antisite pair are bound much stronger in HfC than in TaC. In TaC,
the vacancies are weakly bound into $V_{\alpha}V_{\beta}^{n}$ clusters
when $n\leq3$ and unbound (positive binding energy) in $V_{\alpha}V_{\beta}^{4}$,
$V_{\alpha}V_{\beta}^{5}$ and $V_{\alpha}V_{\beta}^{6}$. The most
stable vacancy cluster is $V_{\alpha}V_{\beta}^{2}$ with the linear
configuration (binding energy $-0.26$ eV). In other words, a metallic
vacancy in TaC is most likely bound to two carbon vacancies but this
binding is relatively weak. The same conclusion was previously reached
in Refs.~\citep{Razumovskiy:2013uk,Razumovskiy:2015wx}. In contrast,
in HfC, the vacancy binding into $V_{\alpha}V_{\beta}^{n}$ clusters
is strong and increases in magnitude with $n$, reaching the most
negative value of $-5.49$ eV in $V_{\alpha}V_{\beta}^{6}$. Thus,
the metallic vacancies in HfC are likely to be surrounded by up to
six carbon vacancies as first neighbors. This conclusion is consistent
with previous reports \citep{Razumovskiy:2013uk,Razumovskiy:2015wx}.

\begin{figure}
\noindent \begin{centering}
\includegraphics[width=0.6\textwidth]{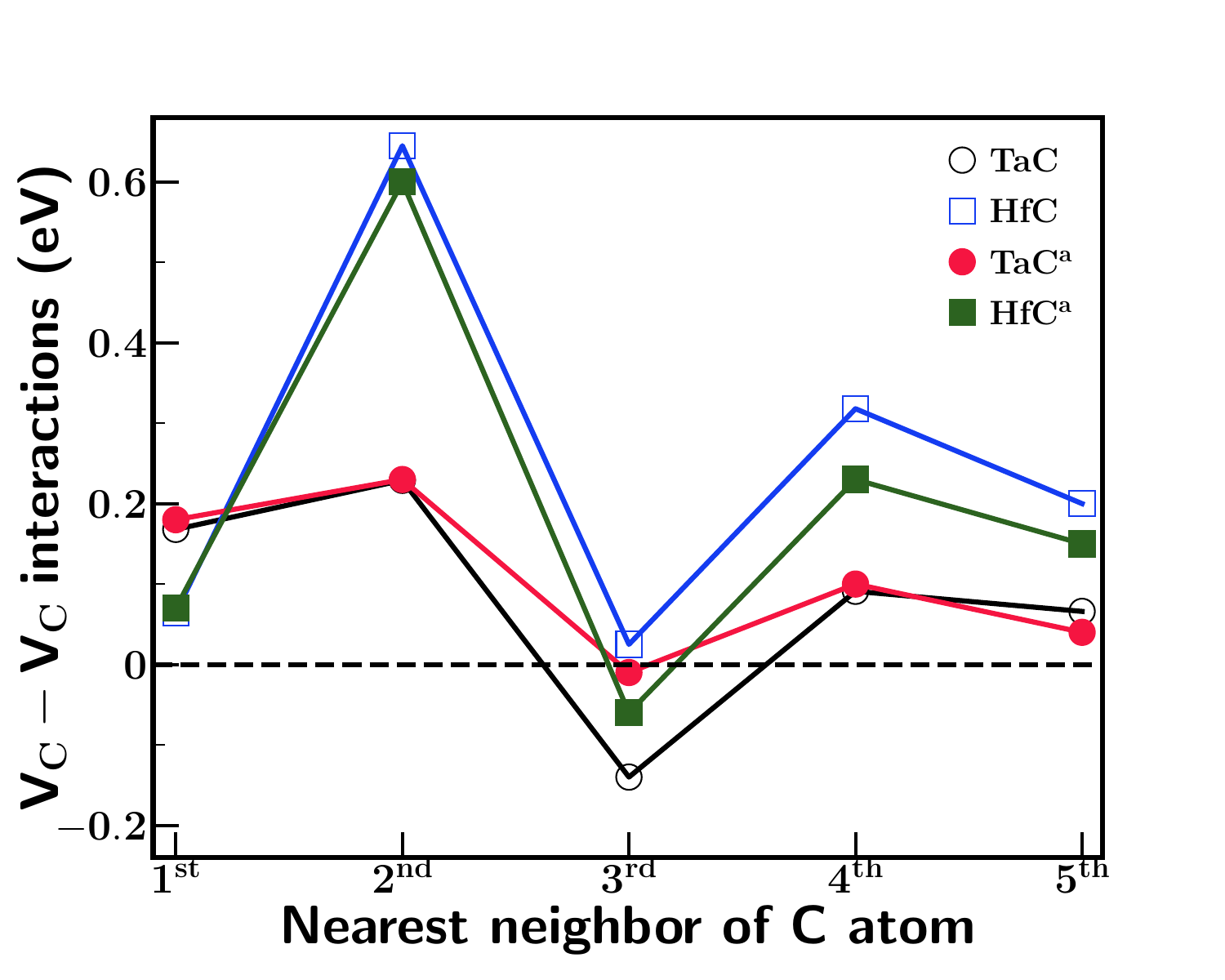} 
\par\end{centering}
\caption{Interaction energy between carbon vacancies in the TaC and HfC carbides
as a function of vacancy separation. The literature data from $^{a}$Ref.~\citep{Salehin:2021ti}
shows similar trends. \label{fig:bining}}
\end{figure}

As discussed by Razumovskiy et al.~\citep{Razumovskiy:2013uk,Razumovskiy:2015wx},
the vacancy binding into $V_{\alpha}V_{\beta}^{n}$ clusters results
from competition between the attraction of carbon vacancies to the
metallic vacancy and their repulsion from each other. Indeed, Fig.~\ref{fig:bining}
shows the interaction energy between carbon vacancies as a function
of separation. In both carbides, the interaction between the first
and second neighbors is repulsive, and the repulsion between second
neighbors is much stronger in HfC than in TaC. Nevertheless, this
repulsion is overpowered by the attraction to the metallic vacancy,
resulting in the strongly bound $V_{\alpha}V_{\beta}^{n}$ clusters
in HfC.

It was shown \citep{Salehin:2021ti} that in the $V_{\alpha}V_{\beta}^{n}$
clusters, a nearby C atom can jump into an interstitial position and
leave a new carbon vacancy behind. In other words, a Frenkel pair
$V_{\beta}I_{B}$ can form next to the $V_{\alpha}V_{\beta}^{n}$
cluster. In TaC, the energy increases in this process, indicating
that the new Frenkel pair is energetically unfavorable. But in HfC,
the energy decreases for certain Frenkel pair orientations. Thus,
the $V_{\alpha}V_{\beta}^{n}$ clusters in HfC are unstable with respect
to Frenkel pair formation in their vicinity. Further investigation
of the $V_{\alpha}V_{\beta}^{n}-I_{B}$ structures in HfC was not
pursued in this work.

\section{Point-defect concentrations \label{sec:Point-defect-concentrations}}

This section discusses a statistical model of point defects in binary
carbides with the B1 structure. As before, we consider a general carbide
AB, where element A is a transition metal and element B is carbon.
Off-stoichiometric carbides are described by the formula A$_{1+x}$B$_{1-x}$,
where $x$ measures the deviation from the perfect stoichiometry.
We consider slight deviations accommodated by small concentrations
of point defects. Under this assumption, interactions among the point
defects can be neglected.

Several methods were proposed for calculating the equilibrium point-defect
concentrations in ordered compounds \citep{Mishin_97a,Mishin_97b,Mishin00b,Lozovoi03b,Woodward:2001aa}.
Here, we follow the quasi-chemical method \citep{Mishin_97a,Mishin00b},
which gives the same results as all other methods but is more straightforward
and transparent. The method was previously applied to compounds dominated
by antisite disorder \citep{Mishin_97a,Mishin00b} and vacancy disorder
\citep{Razumovskiy:2015wx}. Here, we provide a general treatment
without the presumption of a particular disorder mechanism.

The point defects are treated as an ideal gas mixture of several ``chemical
components'' representing the different types of elementary point
defects or their clusters. The equilibrium defect concentrations at
a temperature $T$ are calculated from two conditions: 
\begin{itemize}
\item Equilibrium with respect to chemical reactions in the gas mixture 
\item Material balance preserving the given off-stoichiometry $x$ 
\end{itemize}
If the model considers $M$ defect types, then $(M-1)$ independent
reactions must be chosen, with the material balance condition providing
another equation.

The defects are assigned the chemical potentials 
\begin{equation}
\mu_{d}=\varepsilon_{d}+k_{B}T\ln\dfrac{X_{d}}{\sigma_{d}},\label{eq:2}
\end{equation}
while the formula unit AB is assigned the chemical potential $\mu_{AB}=2\varepsilon_{0}$
($k_{B}$ is Boltzmann's constant). For single vacancies and antisite
defects, $X_{d}$ denotes their fraction of the respective sublattice.
For single interstitial atoms, $X_{d}$ is the occupied fraction on
the available interstitial positions (interstices). Each defect cluster
is assumed to contain at least one vacancy or one antisite. Let us
call this vacancy/antisite the cluster center. The configurational
entropy of the cluster considers all possible locations of its center
on the respective sublattice. The cluster concentration is then defined
as the fraction $X_{d}$ of the sublattice sites occupied by the cluster
centers. In addition, for a given center location, the relaxed cluster
structure can have $\sigma_{d}$ different orientations relative to
the lattice with the same energy. This additional degeneracy of the
micro-states contributes the orientational entropy $k_{B}\ln\sigma_{d}$
per cluster. This explains the appearance of the symmetry factor $\sigma_{d}$
in Eq.(\ref{eq:2}). For example, the vacancy-antisite pair $V_{\beta}B_{\alpha}$
in HfC shown in Fig.~\ref{fig:relaxed_structures}(a) can have six
different orientations with equal energy, thus $\sigma_{d}=6$. In
this case, either $V_{\beta}$ or $B_{\alpha}$ can be taken as the
cluster center. For elementary defects $\sigma_{d}=1$. The symmetry
factors for all defect clusters considered in this work are summarized
in Table \ref{table:bindingenergy}.

Although the chemical potentials of the defects are expressed through
the ``raw'' energies, it can be shown \citep{Mishin00b} that all
reference energies cancel out and do not affect the defect concentrations
predicted by this method. We emphasize that the model only includes
the configurational and orientational micro-states and neglects the
vibrational, electronic, and all other forms of free energy.

Although any choice of the $(M-1)$ reactions is equally legitimate,
we find the following set of reactions most intuitive. For reactions
among the six elemental defects, we choose 
\begin{equation}
V_{\alpha}+V_{\beta}=-AB,\label{eq:reaction_1}
\end{equation}
\begin{equation}
B_{\alpha}+A_{\beta}=0,\label{eq:reaction_2}
\end{equation}

\begin{equation}
V_{\alpha}=B_{\alpha}+V_{\beta},\label{eq:reaction_3}
\end{equation}

\begin{equation}
V_{\alpha}+I_{A}=0,\label{eq:reaction_4}
\end{equation}

\begin{equation}
V_{\beta}+I_{B}=0.\label{eq:reaction_5}
\end{equation}
In reaction (\ref{eq:reaction_1}), a vacancy pair is created by removing
one formula unit of the carbide. In reaction (\ref{eq:reaction_2}),
a pair of antisite defects is created without adding or removing atoms.
In reaction (\ref{eq:reaction_3}), an atom B fills the vacancy $V_{\alpha}$,
creating an antisite defect $B_{\alpha}$ and leaving a vacancy $V_{\beta}$
behind. Again, the system remains closed. Finally, reactions (\ref{eq:reaction_4})
and (\ref{eq:reaction_5}) describe the Frenkel pair formation by
atoms A and B, respectively. For every cluster $d$ composed of elementary
defects $d_{i}$, we write the dissociation-recombination reaction
\begin{equation}
d=\sum_{i}d_{i}.\label{eq:reaction_6}
\end{equation}
It is assumed that the point defects participating in the reactions
are separated well enough to neglect their interaction.

The equations describing dynamic equilibrium with respect to the defect
reactions are obtained by replacing the defect symbols by the respective
chemical potentials. The equations obtained have the form of the mass
action law known from chemistry. In particular, reactions (\ref{eq:reaction_1})-(\ref{eq:reaction_5})
yield the equations 
\begin{equation}
X_{V_{\alpha}}X_{V_{\beta}}=\exp\left(-\dfrac{\varepsilon_{V_{\alpha}}+\varepsilon_{V_{\beta}}+2\varepsilon_{0}}{k_{B}T}\right),\label{eq:mass_action_1}
\end{equation}
\begin{equation}
X_{B_{\alpha}}X_{A_{\beta}}=\exp\left(-\dfrac{\varepsilon_{B_{\alpha}}+\varepsilon_{A_{\beta}}}{k_{B}T}\right),\label{eq:mass_action_2}
\end{equation}
\begin{equation}
\dfrac{X_{V_{\alpha}}}{X_{B_{\alpha}}X_{V_{\beta}}}=\exp\left(-\dfrac{\varepsilon_{V_{\alpha}}-\varepsilon_{B_{\alpha}}-\varepsilon_{V_{\beta}}}{k_{B}T}\right),\label{eq:mass_action_3}
\end{equation}

\begin{equation}
X_{V_{\alpha}}X_{I_{A}}=\exp\left(-\dfrac{\varepsilon_{V_{\alpha}}+\varepsilon_{I_{A}}}{k_{B}T}\right),\label{eq:mass_action_4}
\end{equation}

\begin{equation}
X_{V_{\beta}}X_{I_{B}}=\exp\left(-\dfrac{\varepsilon_{V_{\beta}}+\varepsilon_{I_{B}}}{k_{B}T}\right).\label{eq:mass_action_5}
\end{equation}
Similarly, each cluster dissociation-recombination reaction (\ref{eq:reaction_6})
gives the equation 
\begin{equation}
\dfrac{X_{d}}{\prod_{i}X_{d_{i}}}=\sigma_{d}\exp\left(-\dfrac{\varepsilon_{d}-\sum_{i}\varepsilon_{d_{i}}}{k_{B}T}\right),\label{eq:mass_action 6}
\end{equation}
where the numerator $\varepsilon_{d}-\sum_{i}\varepsilon_{d_{i}}$
in the right-hand side has the meaning of the binding energy of the
cluster. For example, for a vacancy cluster $V_{\alpha}V_{\beta}^{n}$
we have 
\begin{equation}
\dfrac{X_{V_{\alpha}V_{\beta}^{n}}}{X_{V_{\alpha}}\left(X_{V_{\beta}}\right)^{n}}=\sigma_{V_{\alpha}V_{\beta}^{n}}\exp\left(-\dfrac{\varepsilon_{V_{\alpha}V_{\beta}^{n}}-\varepsilon_{V_{\alpha}}-n\varepsilon_{V_{\beta}}}{k_{B}T}\right).\label{eq:mass_action 7}
\end{equation}
It is easy to see that the number of equations obtained is $(M-1)$.

The mass balance equation is generally nonlinear \citep{Mishin00b}
with respect to the defect concentrations. However, a linear approximation
can be applied considering that the defect concentrations are small.
This approximation neglects all terms quadratic in the defects concentrations,
i.e., terms representing products of different concentrations or their
squares. The following linear equation can be derived: 
\begin{eqnarray}
x & = & \dfrac{1}{4}\left(X_{V_{\beta}}-X_{V_{\alpha}}\right)+\dfrac{1}{2}\left(X_{A_{\beta}}-X_{B_{\alpha}}\right)-\dfrac{1}{2}\nu\left(X_{I_{B}}-X_{I_{A}}\right)\nonumber \\
 & + & \dfrac{1}{4}\sum_{d}X_{d}\left[L_{V_{\beta}}^{d}-L_{V_{\alpha}}^{d}+2\left(L_{A_{\beta}}^{d}-L_{B_{\alpha}}^{d}\right)+\left(L_{I_{B}}^{d}-L_{I_{A}}^{d}\right)\right].\label{eq:balance}
\end{eqnarray}
Here, the first line represents the deviation from the stoichiometry
due to the elementary defects, $\nu$ being the number of interstitial
positions per lattice site. We only consider tetrahedral interstitials,
for which $\nu=8$. The second line is the contribution of the defect
clusters $d$, where $L_{d_{i}}^{d}$ is the number of elementary
defects $d_{i}$ in the cluster $d$. For example, for the vacancy
cluster $V_{\alpha}V_{\beta}^{n}$ we have $L_{V_{\alpha}}^{V_{\alpha}V_{\beta}^{n}}=1$,
$L_{V_{\beta}}^{V_{\alpha}V_{\beta}^{n}}=n$, and $L_{B_{\alpha}}^{V_{\alpha}V_{\beta}^{n}}=L_{A_{\beta}}^{V_{\alpha}V_{\beta}^{n}}=0$.
The complete list of the $L-$numbers is provided in the Appendix,
along with Eq.(\ref{eq:balance}) specialized for the chosen set of
defect clusters.

Note that the divacancy, the antisite pair, and the Frenkel pairs
do not affect the material balance. Accordingly, the terms representing
these pairs mutually cancel in Eq.(\ref{eq:balance}). This is true
for any composition-conserving defect cluster. The formation of such
clusters is accompanied by addition or removal of $m$ formula units
AB or does not require any addition or removal ($m=0$). It is easy
to show that the concentration of composition-conserving clusters
is 
\begin{equation}
X_{d}=\sigma_{d}\exp\left(-\dfrac{\varepsilon_{d}+2m\varepsilon_{0}}{k_{B}T}\right).\label{eq:clusters}
\end{equation}
For example, $m=1$ for divacancies and $m=0$ for antisite pairs
and Frenkel pairs. This concentration is independent of the off-stoichiometry
and can be immediately calculated at any given temperature without
solving any equations.

Eqs.(\ref{eq:mass_action_1})-(\ref{eq:mass_action 6}) and (\ref{eq:balance})
(less the equations for the composition-conserving clusters) constitute
a complete set of equations that must be solved for the defect concentrations
numerically.

\begin{figure}
(a)\includegraphics[width=0.47\textwidth]{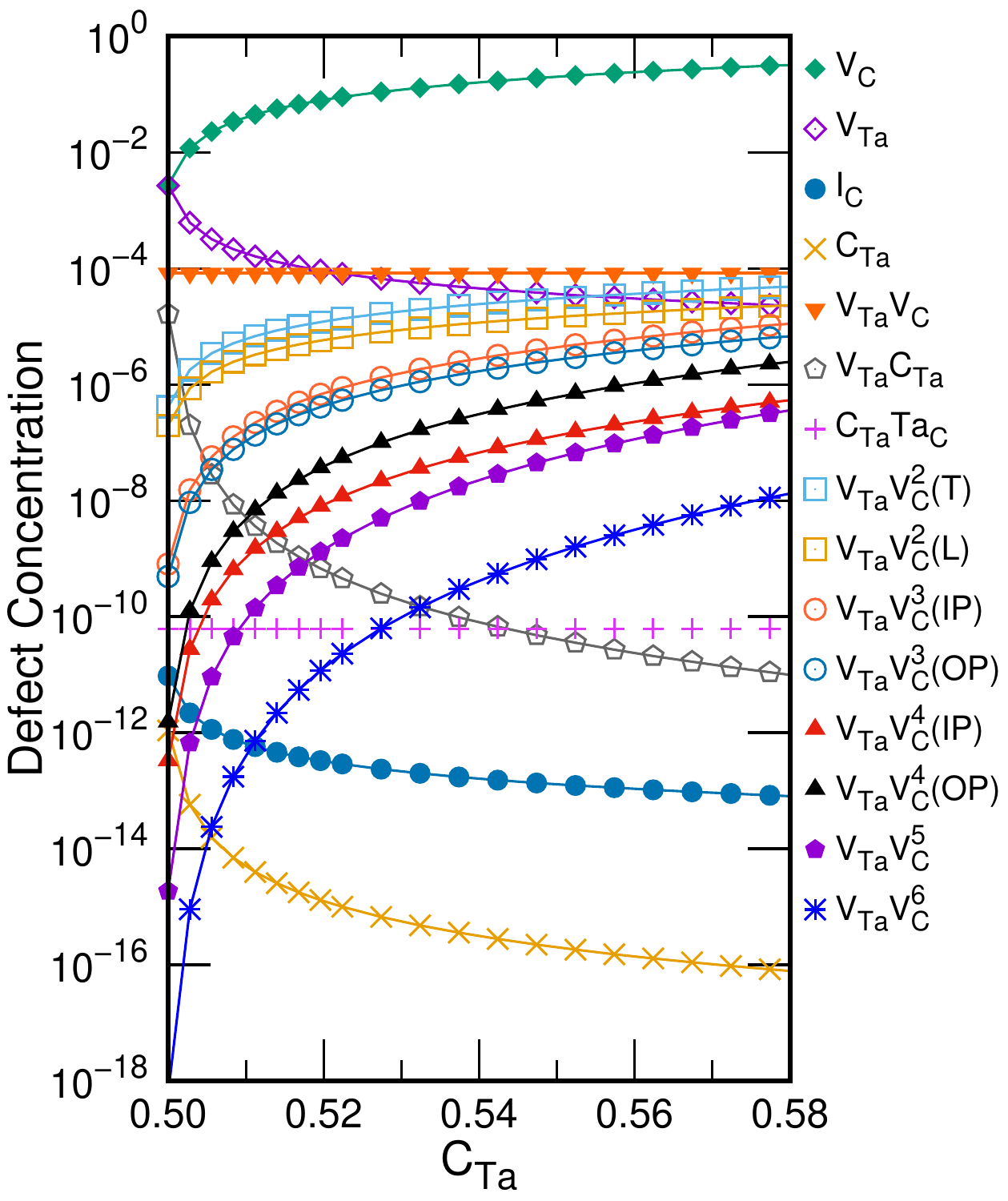}(b)\includegraphics[width=0.47\textwidth]{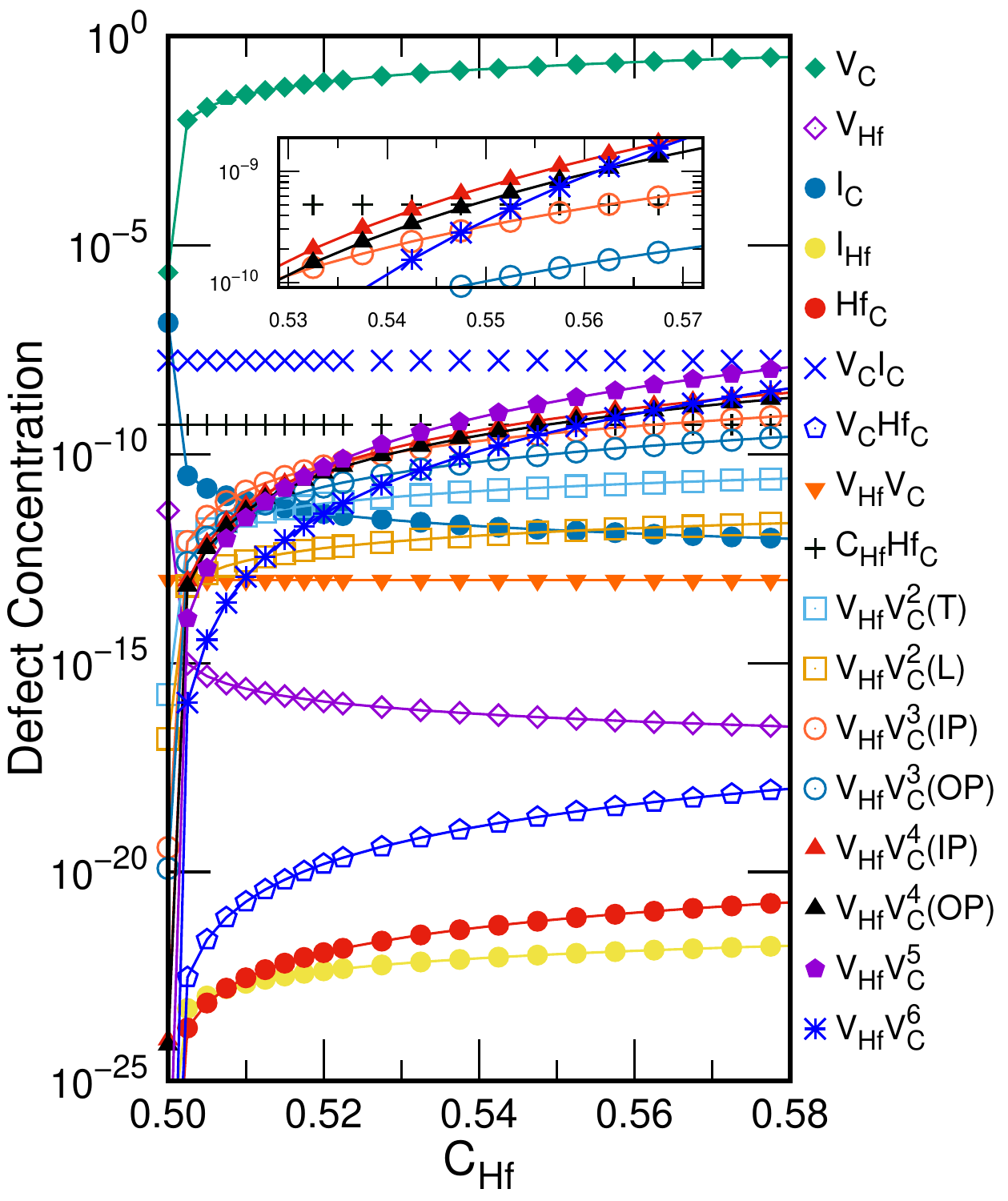}

\caption{Composition dependence of point defect concentrations in (a) TaC and
(b) HfC at 2500 K.\label{fig:Composition-dependence}}
\end{figure}

Figures \ref{fig:Composition-dependence}(a) and (b) present the computed
composition dependencies of the defect concentrations in TaC and HfC,
respectively. The temperature is fixed at 2500 K, but results for
other temperatures are qualitatively similar. The Supplementary Information
file shows the plots for 3500 K. The plots in Fig.~\ref{fig:Composition-dependence}
only include the stoichiometric and metal-rich compositions ($x\geq0$)
because carbon-rich compositions ($x<0$) are unstable according to
the phase diagram \citep{Mehdikhan:2017tj,Shabalin:2019ws}. In the
plots, the chemical compositions are measured by the fraction of metallic
atoms $c_{A}=1+x$ (A = Ta or Hf).

From Figs.~\ref{fig:Composition-dependence}(a) and (b), it is evident
that thermal disorder in the stoichiometric carbides is dominated
by Ta and C vacancies in TaC and by carbon vacancies and carbon interstitials
in HfC. This difference reflects the different energetics of the point
defects and, ultimately, the difference in the chemical bonding in
the two carbides. The Hf vacancy concentration in stoichiometric HfC
is extremely small relative to the carbon defects. As the off-stoichiometry
increases, so does the carbon vacancy concentration, showing that
the carbon vacancies are the constitutional defects on the metal-rich
side in both carbides. As expected from the mass action law, the carbon
vacancies suppress the metallic vacancies and carbon interstitials.
In off-stoichiometric Ta$_{1+x}$C$_{1-x}$ with $x<0.04$, the metallic
sublattice is dominated by single vacancies $V_{\mathrm{Ta}}$. At
larger deviations from the stoichiometry ($x>0.04$), the Ta vacancies
start forming divacancies $V_{\mathrm{Ta}}V_{\mathrm{C}}$ and tri-vacancies
$V_{\mathrm{Ta}}V_{\mathrm{C}}^{2}$, whose concentrations eventually
become comparable to that of single vacancies. In off-stoichiometric
Hf$_{1+x}$C$_{1-x}$, the dominant metallic defects likewise depend
on the chemical composition. At small deviations from the stoichiometry
($x<0.001$), the leading metallic defects are single vacancies $V_{\mathrm{Hf}}$.
As the off-stoichiometry increases, the concentrations of vacancy
clusters $V_{\mathrm{Hf}}V_{\mathrm{C}}^{n}$ rapidly grow and at
$x>0.001$ they exceed the $V_{\mathrm{Hf}}$ concentration. The six-vacancy
clusters $V_{\mathrm{Hf}}V_{\mathrm{C}}^{5}$ have the highest concentration,
although the clusters with smaller $n$ come close.

\section{Effective defect formation energies}

\begin{figure}
\begin{centering}
(a)\includegraphics[width=0.54\textwidth]{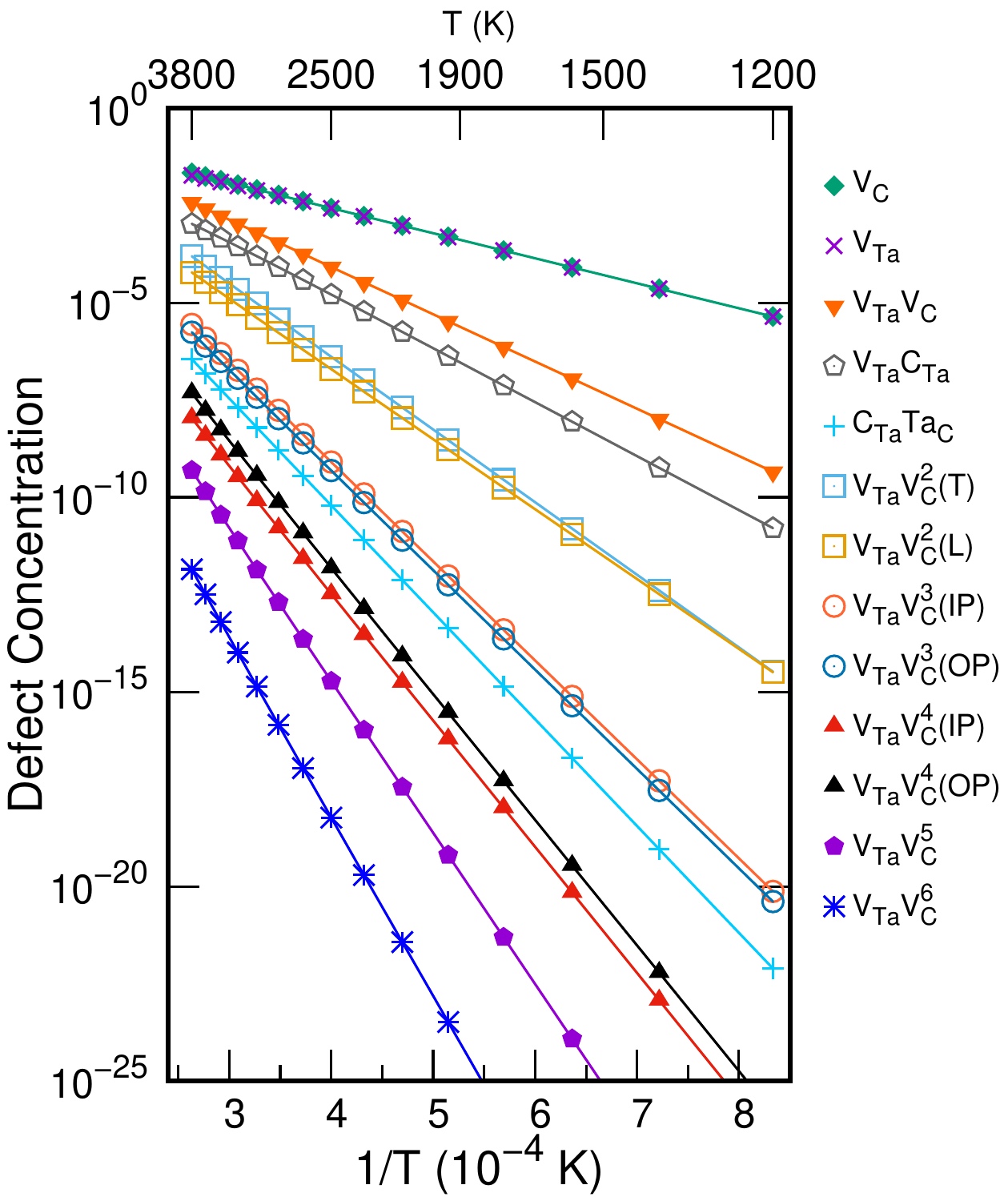}
\par\end{centering}
\begin{centering}
\bigskip{}
\par\end{centering}
\begin{centering}
(b)\includegraphics[width=0.54\textwidth]{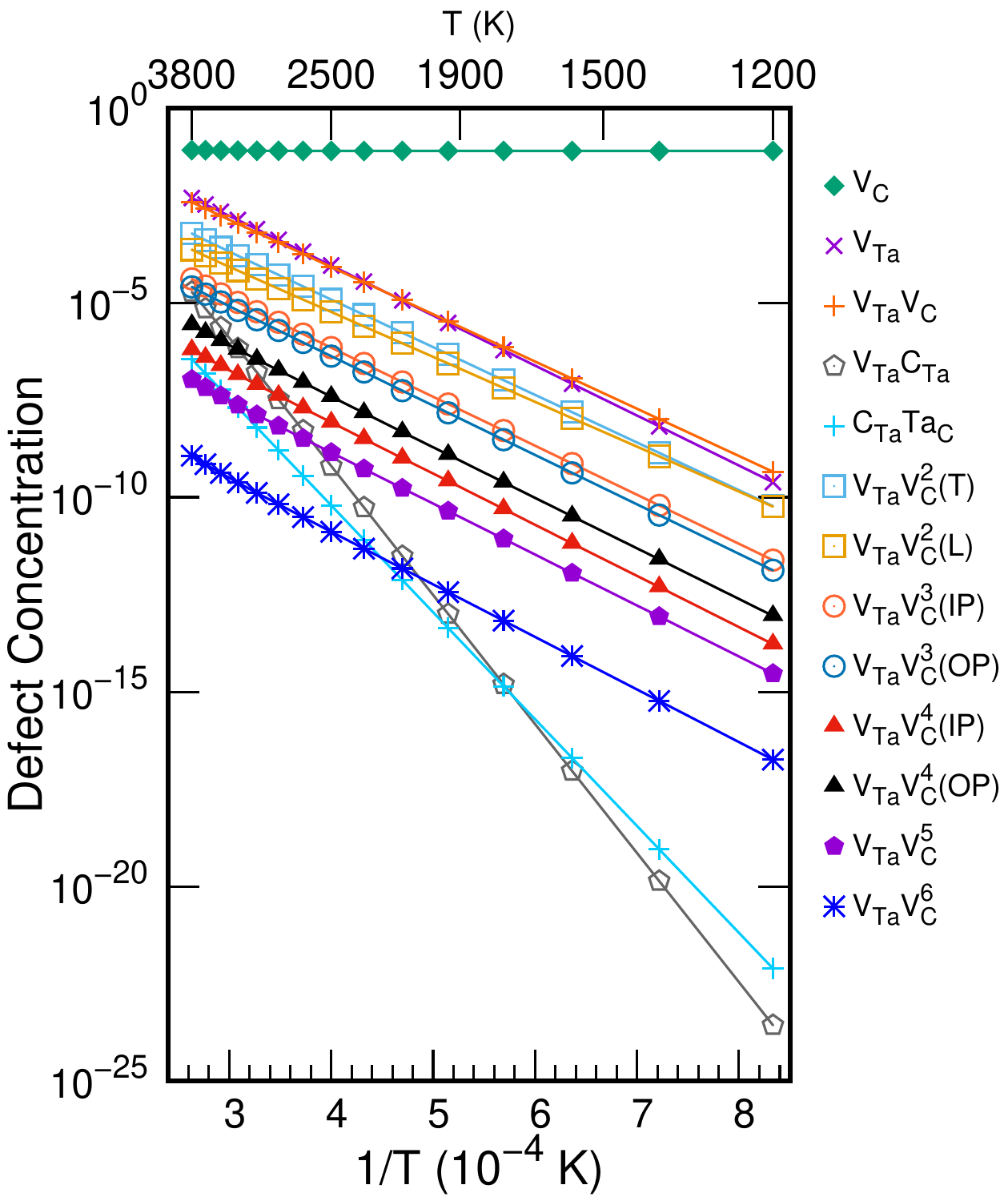}
\par\end{centering}
\caption{Arrhenius plots of point defect concentrations in TaC at (a) stoichiometric
composition and (b) Ta-rich composition of $c_{\mathrm{Ta}}=0.52$
($x=0.02$).\label{fig:Arrhenius-TaC}}
\end{figure}

\begin{figure}
\begin{centering}
(a)\includegraphics[width=0.54\textwidth]{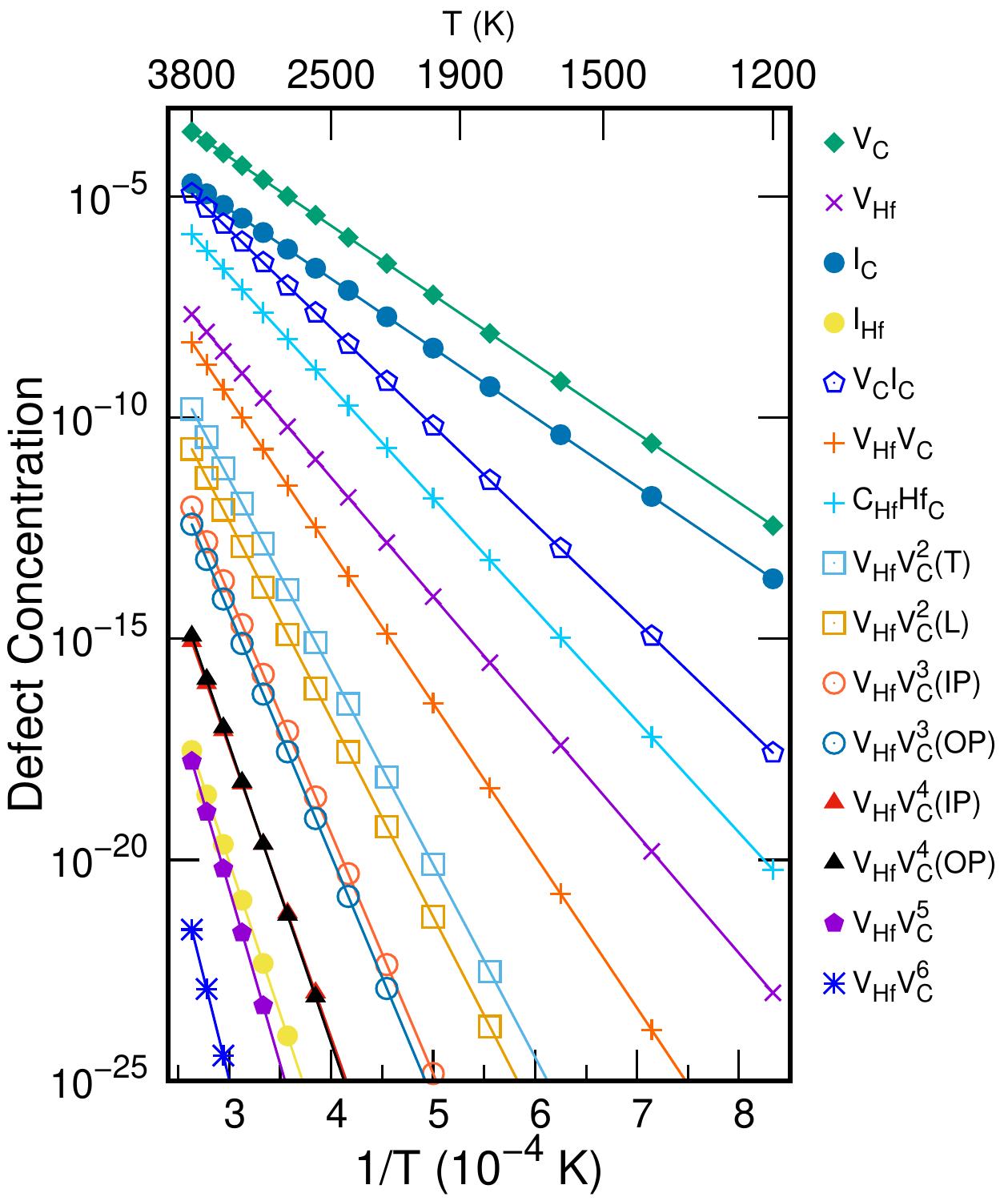}
\par\end{centering}
\begin{centering}
\bigskip{}
\par\end{centering}
\begin{centering}
(b)\includegraphics[width=0.54\textwidth]{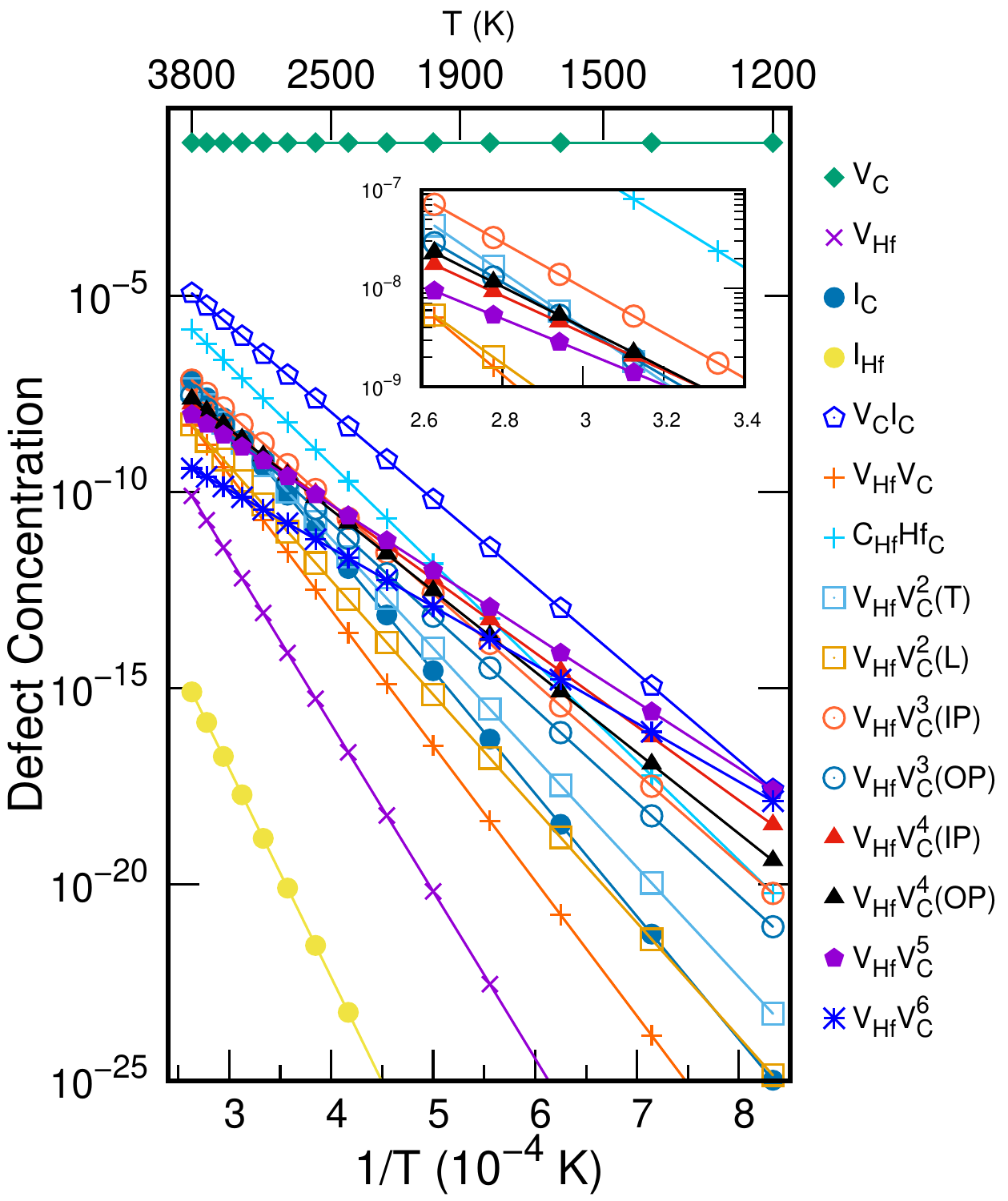}
\par\end{centering}
\caption{Arrhenius plots of point defect concentrations in HfC at (a) stoichiometric
composition and (b) Hf-rich composition of $c_{\mathrm{Hf}}=0.52$
($x=0.02$).\label{fig:Arrhenius-HfC}}
\end{figure}

Figures \ref{fig:Arrhenius-TaC} and \ref{fig:Arrhenius-HfC} present
the Arrhenius diagrams, $\log X_{d}$ versus $1/T$, of the defect
concentrations in TaC and HfC, respectively. The chemical compositions
are fixed at $x=0$ (stoichiometry) and $x=0.02$ (representative
off-stoichiometric state). Recall that the composition-conserving
defect clusters strictly follow the Arrhenius law, see Eq.(\ref{eq:clusters}).
As a result, their plots are represented by perfect straight lines
on the Arrhenius diagrams and perfect horizontal lines in the composition
plots (Fig.~\ref{fig:Composition-dependence}). The single defects
and non-conserving defect clusters do not follow the Arrhenius law
exactly but their plots still look fairly straight. As a good approximation,
all defect concentrations can be represented in the Arrhenius form
\begin{equation}
X_{d}=X_{d}^{0}\exp\left(-\dfrac{\overline{\varepsilon}_{d}}{k_{B}T}\right)\label{eq:clusters-1}
\end{equation}
with appropriate prefactors $X_{d}^{0}$ and effective formation energies
$\overline{\varepsilon}_{d}$. Although not exact, Eq.(\ref{eq:clusters-1})
is useful for applications and comparison with experiments.

As discussed in Ref.~\citep{Mishin00b}, Eq.(\ref{eq:clusters-1})
works best in two limiting cases: when one defect concentration is
much higher than all other concentrations and when the compound is
stoichiometric with thermal disorder strongly dominated by two defect
types. In the first case, the right-hand side of Eq.(\ref{eq:balance})
can be represented by a single term, so the respective defect concentration
is proportional to $x$. In the second case, the right-hand side of
Eq.(\ref{eq:balance}) is represented by two terms while the left-hand
side is zero; thus the two defect concentrations are equal up to a
numerical factor. In both cases, the material balance equation is
simplified, and Eqs.(\ref{eq:mass_action_1})-(\ref{eq:mass_action 6})
can be solved analytically with solutions in the form of Eq.(\ref{eq:clusters-1}).

The first case is realized in the off-stoichiometric carbides dominated
by constitutional carbon vacancies. In our notation, $X_{V_{\beta}}$
is much greater than all other concentrations $X_{d}$. Accordingly,
the balance equation (\ref{eq:balance}) is simplified to $x=\frac{1}{4}X_{V_{\beta}}$,
from which $X_{V_{\beta}}=4x$. Thus, $X_{V_{\beta}}$ is temperature-independent
and proportional to the off-stoichiometry parameter $x$. Inserting
this $X_{V_{\beta}}$ in Eq.(\ref{eq:mass_action_1}), we obtain the
metal vacancy concentration 
\begin{equation}
X_{V_{\alpha}}=\dfrac{1}{4x}\exp\left(-\dfrac{\varepsilon_{V_{\alpha}}+\varepsilon_{V_{\beta}}+2\varepsilon_{0}}{k_{B}T}\right).\label{eq:18}
\end{equation}
Next, we find the concentrations of the vacancy clusters $V_{\alpha}V_{\beta}^{n}$
using Eq.(\ref{eq:mass_action 7}): 
\begin{equation}
X_{V_{\alpha}V_{\beta}^{n}}=\sigma_{V_{\alpha}V_{\beta}^{n}}(4x)^{n-1}\exp\left(-\dfrac{\varepsilon_{V_{\alpha}V_{\beta}^{n}}-(n-1)\varepsilon_{V_{\beta}}+2\varepsilon_{0}}{k_{B}T}\right).\label{eq:19}
\end{equation}
For the interstitial concentrations we use Eqs.(\ref{eq:mass_action_4})
and (\ref{eq:mass_action_5}) to obtain 
\begin{equation}
X_{I_{A}}=4x\exp\left(-\dfrac{\varepsilon_{I_{A}}-\varepsilon_{V_{\beta}}-2\varepsilon_{0}}{k_{B}T}\right).\label{eq:20}
\end{equation}
\begin{equation}
X_{I_{B}}=\dfrac{1}{4x}\exp\left(-\dfrac{\varepsilon_{I_{B}}+\varepsilon_{V_{\beta}}}{k_{B}T}\right).\label{eq:21}
\end{equation}
This chain of calculations can be continued to obtain all other defect
concentrations. Note that they all have the Arrhenius form (\ref{eq:clusters-1}).
The respective effective formation energies and prefactors are summarized
in Table \ref{tab:Arrhenius_eqns_x}. Equations (\ref{eq:18}) and
(\ref{eq:21}) confirm the trend mentioned above: deviations from
stoichiometry with $x>0$ suppress the metallic vacancies and carbon
interstitials. At the same time, such deviations promote the formation
of the vacancy clusters $V_{\alpha}V_{\beta}^{n}$ and metallic interstitials
$I_{A}$.

The second case is realized in stoichiometric carbides. The leading
thermal defects in the two carbides are different. In TaC, such defects
are the vacancies $V_{\alpha}$ and $V_{\beta}$. The balance equation
(\ref{eq:balance}) gives $V_{\alpha}=V_{\beta}$, which we combine
with Eq.(\ref{eq:mass_action_1}) to obtain 
\begin{equation}
X_{V_{\alpha}}=X_{V_{\beta}}=\exp\left(-\dfrac{\varepsilon_{V_{\alpha}}+\varepsilon_{V_{\beta}}+2\varepsilon_{0}}{2k_{B}T}\right).\label{eq:22}
\end{equation}
The interstitial concentrations are then obtained by inserting Eq.(\ref{eq:22})
into Eqs.(\ref{eq:mass_action_4}) and (\ref{eq:mass_action_5}):
\begin{equation}
X_{I_{A}}=\exp\left(-\dfrac{2\varepsilon_{I_{A}}-2\varepsilon_{0}+\varepsilon_{V_{\alpha}}-\varepsilon_{V_{\beta}}}{2k_{B}T}\right),\label{eq:23}
\end{equation}
\begin{equation}
X_{I_{B}}=\exp\left(-\dfrac{2\varepsilon_{I_{B}}-2\varepsilon_{0}+\varepsilon_{V_{\beta}}-\varepsilon_{V_{\alpha}}}{2k_{B}T}\right).\label{eq:24}
\end{equation}
The concentrations of the vacancy clusters $V_{\alpha}V_{\beta}^{n}$
are readily calculated by inserting the single vacancy concentrations
from Eq.(\ref{eq:22}) into Eq.(\ref{eq:mass_action 7}), which gives
\begin{equation}
X_{V_{\alpha}V_{\beta}^{n}}=\sigma_{V_{\alpha}V_{\beta}^{n}}\exp\left(-\dfrac{2\varepsilon_{V_{\alpha}V_{\beta}^{n}}+(n-1)(\varepsilon_{V_{\alpha}}-\varepsilon_{V_{\beta}})+2(n+1)\varepsilon_{0}}{2k_{B}T}\right).\label{eq:25}
\end{equation}
The remaining defect concentrations are calculated similarly, and
the Arrhenius parameters obtained are summarized in Table \ref{tab:Arrhenius_eqns_x0}.

In stoichiometric HfC, the leading thermal defects are carbon vacancies
$V_{\beta}$ and carbon interstitials $I_{B}$. The balance equation
(\ref{eq:balance}) gives $X_{V_{\beta}}=2\nu X_{I_{B}}$ (recall
that $\nu=8$ in the B1 structure). Using Eq.(\ref{eq:mass_action_5})
we have 
\begin{equation}
X_{V_{\beta}}=\sqrt{2\nu}\exp\left(-\dfrac{\varepsilon_{I_{B}}+\varepsilon_{V_{\beta}}}{2k_{B}T}\right).\label{eq:26}
\end{equation}
\begin{equation}
X_{I_{B}}=\dfrac{1}{\sqrt{2\nu}}\exp\left(-\dfrac{\varepsilon_{I_{B}}+\varepsilon_{V_{\beta}}}{2k_{B}T}\right).\label{eq:27}
\end{equation}
Next, the metal vacancy concentration is obtained from Eq.(\ref{eq:mass_action_1}):
\begin{equation}
X_{V_{\alpha}}=\dfrac{1}{\sqrt{2\nu}}\exp\left(-\dfrac{2\varepsilon_{V_{\alpha}}+\varepsilon_{V_{\beta}}-\varepsilon_{I_{B}}+4\varepsilon_{0}}{2k_{B}T}\right),\label{eq:28}
\end{equation}
while Eq.(\ref{eq:mass_action_4}) gives 
\begin{equation}
X_{I_{A}}=\sqrt{2\nu}\exp\left(-\dfrac{2\varepsilon_{I_{A}}-\varepsilon_{V_{\beta}}+\varepsilon_{I_{B}}-4\varepsilon_{0}}{2k_{B}T}\right).\label{eq:29}
\end{equation}
The vacancy cluster concentrations are obtained from Eq.(\ref{eq:mass_action 7}),
which gives 
\begin{equation}
X_{V_{\alpha}V_{\beta}^{n}}=\sigma_{V_{\alpha}V_{\beta}^{n}}(2\nu)^{(n-1)/2}\exp\left(-\dfrac{2\varepsilon_{V_{\alpha}V_{\beta}^{n}}+(n-1)(\varepsilon_{I_{B}}-\varepsilon_{V_{\beta}})+4\varepsilon_{0}}{2k_{B}T}\right).\label{eq:30}
\end{equation}
Continuing the calculations, we derive the Arrhenius parameters of
all other defects summarized in Table \ref{tab:Arrhenius_eqns_x0}.

Table \ref{tab:Arrhenius-parameters} reports the numerical values
of the Arrhenius parameters of the point defects in TaC and HfC obtained
from the first-principles raw energies. Inserting these values in
Eq.(\ref{eq:clusters-1}), the Arrhenius lines obtained closely approximate
the numerical solutions shown in Figs.~\ref{fig:Arrhenius-TaC} and
\ref{fig:Arrhenius-HfC}.

\section{Discussion \label{sec:Discussion}}

The goal of this work was to understand and predict the point defects
in the binary carbides TaC and HfC by DFT calculations. To this end,
a statistical-mechanical model has been developed capable of predicting
point-defect concentrations in binary compounds with the AB stoichiometry.
The model applies to both single defects and defect clusters of any
complexity. Although our primary interest is in the particular carbides
TaC and HfC, the the model is general enough to be applied to other
binary carbides, nitrides, and borides with the AB stoichiometry.
It can also be generalized to ordered compounds with different crystal
structures and stoichiometries \citep{Mishin00b}. 

The model generalizes the previous treatment of point defects in intermetallic
compounds \citep{Mishin00b}. Thermal and compositional disorder in
intermetallic compounds is also governed by point defects and has
been studied by similar DFT statistical mechanics methods \citep{Mishin_97a,Mishin_97b,Mishin00b,Lozovoi03b}.
However, intermetallics do not usually support interstitial defects,
and vacancies display a weak clustering trend. Thus, the point-defect
system is much simpler than in the carbides.

In the present model, the free energy of the point defects includes
the configurational and orientational effects but neglects other contributions,
such as atomic vibrations. This is not a severe limitation. As discussed
previously \citep{Mishin00b}, the effect of vibrations can be included
by replacing the ``raw'' defect energy $\varepsilon_{d}$ by the
``raw'' free energy $f_{d}=\varepsilon_{d}-Ts_{d}$, where $s_{d}$
is the change in the vibrational entropy due to the defect formation
in a perfect crystal. DFT-based calculations of $s_{d}$ are challenging,
especially for defect clusters, but calculations with interatomic
potentials (traditional or machine-learning type \citep{Mishin-2021})
are straightforward. The effect of applied pressure on the point-defect
concentrations can also be included, provided the defect formation
volumes can be computed.

As mentioned in Section \ref{sec:Introduction}, our primary motivation
for calculating the point-defect concentrations is that they constitute
a required ingredient for diffusion calculations. We will briefly
discuss the possible diffusion mechanisms suggested by the present
results, focusing on the most realistic case of carbon-deficient chemical
compositions Ta$_{1+x}$C$_{1-x}$ ($x>0$). As evident from the calculations,
deviations from the stoichiometry towards carbon-deficient compositions
are accommodated by carbon vacancies. The carbon vacancy concentration
is virtually temperature-independent and can reach a few percent (recall
that $X_{V_{\beta}}\approx4x$), dominating over all other point defects.
The fraction of carbon vacancies bound into clusters is orders of
magnitude smaller. Thus, it is highly probable that carbon diffusion
is mediated by single vacancy jumps on the carbon sublattice. Interactions
among the carbon vacancies can hardly affect their diffusion given
their repulsion as first and second neighbors (Fig.~\ref{fig:bining}).
To a good approximation, carbon diffusion can be treated as occurring
by the simple vacancy mechanism on the FCC sublattice with the geometric
correlation factor. The same reasoning applies to carbon diffusion
in Hf$_{1+x}$C$_{1-x}$ ($x>0$).

Diffusion of the metallic atoms is more complex. In TaC, single Ta
vacancies have a slightly higher concentration than the divacancies
$V_{\mathrm{Ta}}V_{\mathrm{C}}$ and trivacancies $V_{\mathrm{Ta}}V_{\mathrm{C}}^{2}$.
However, the literature data \citep{Yu:2015us,Tang:2020wb} indicate
that Ta vacancies have a lower jump barrier when bound with carbon
vacancies. Thus, divacancies and possibly $V_{\mathrm{Ta}}V_{\mathrm{C}}^{2}$(T)
clusters (triangular configuration) should also be considered. The
linear clusters $V_{\mathrm{Ta}}V_{\mathrm{C}}^{2}$(L) have a lower
concentration and are in a locked configuration: the Ta vacancy cannot
make a jump without breaking away from this cluster. One complication
is that atomic diffusion mediated by divacancies and trivacancies
is accompanied by jump correlation effects, which can be significant.
Approximate semi-analytical methods exist for calculating the jump
correlation factors in the B1 structure under the divacancy mechanism
\citep{Howard-1966,Belova:2009aa}. However, more accurate calculations
utilize kinetic Monte Carlo (KMC) simulations \citep{Belova:2009aa}.
For trivacancies, KMC is the most viable option.

In HfC, Hf diffusion is mediated by the $V_{\mathrm{Hf}}V_{\mathrm{C}}^{n}$
clusters. The size of the dominant cluster depends on temperature
and chemical composition. Fig.~\ref{fig:Arrhenius-HfC} shows that
at the off-stoichiometry of $x=0.02$, the clusters having the highest
concentration are $V_{\mathrm{Hf}}V_{\mathrm{C}}^{3}$(IP) above $\sim2500$
K and $V_{\mathrm{Hf}}V_{\mathrm{C}}^{5}$ below $\sim2500$ K. The
transition temperature depends on the off-stoichiometry. Note, however,
that in both cases, the Hf vacancy cannot make a jump without destroying
the cluster. In $V_{\mathrm{Hf}}V_{\mathrm{C}}^{3}$(IP), such a jump
causes one of the dissociation reactions $V_{\mathrm{Hf}}V_{\mathrm{C}}^{3}\mathrm{(IP)}\rightarrow V_{\mathrm{Hf}}V_{\mathrm{C}}^{2}\mathrm{(T)}+V_{\mathrm{C}}$
or $V_{\mathrm{Hf}}V_{\mathrm{C}}^{3}\mathrm{(IP)}\rightarrow V_{\mathrm{Hf}}V_{\mathrm{C}}+2V_{\mathrm{C}}$.
The isolated carbon vacancies produced by these reactions repel each
other and are unlikely to recombine into the original cluster. In
$V_{\mathrm{Hf}}V_{\mathrm{C}}^{5}$, a Hf vacancy jump causes one
of the dissociations $V_{\mathrm{Hf}}V_{\mathrm{C}}^{5}\rightarrow V_{\mathrm{Hf}}+5V_{\mathrm{C}}$,
$V_{\mathrm{Hf}}V_{\mathrm{C}}^{5}\rightarrow V_{\mathrm{Hf}}V_{\mathrm{C}}+4V_{\mathrm{C}}$,
or $V_{\mathrm{Hf}}V_{\mathrm{C}}^{5}\rightarrow V_{\mathrm{Hf}}V_{\mathrm{C}}^{2}\mathrm{(T)}+3V_{\mathrm{C}}$,
which again produces isolated carbon vacancies repelling each other.
This behavior reflects the general trend that vacancy cluster migration
by single jumps is subject to severe geometric constraints and strong
correlation effects.

Three alternatives can be considered. One is to allow the vacancy
clusters to evolve by a chain of dissociation-recombination reactions
among all possible clusters \citep{Tang:2020wb}. One should then
resort to KMC simulations as analytical treatment of the atomic transport
caused by such chains of reactions could be impractical. A second
approach is to consider collective (simultaneous) jumps of a group
of atoms filling the vacant sites and shifting the entire vacancy
cluster to a new position. Such mechanisms were discussed in Ref.~\citep{Razumovskiy:2013uk}
for the TiC and ZrC carbides and seem plausible. For simple clusters
such as divacancies, calculating the minimum-energy path is straightforward,
but multi-vacancy clusters present a challenge. One can test a set
of \emph{a priori} chosen atomic trajectories and select one with
the lowest barrier \citep{Razumovskiy:2013uk}. A more general treatment
should use a saddle-point search algorithm not relying on \emph{a
priori} assumptions about which atom will land in which position.
Yet another approach is to use molecular dynamics to discover the
migration mechanisms. This approach is appealing but will likely require
a surrogate model such as a machine-learning potential. It should
also be noted that comparison of different vacancy clusters cannot
be made solely from their equilibrium concentration and migration
barrier. The attempt frequency also matters. The latter depends on
the total mass $m$ of the collectively jumping atoms (approximately
as $1/\sqrt{m}$) and cluster-specific vibrational modes.

\section{Conclusions}

We have combined DFT calculations with a statistical-mechanical model
to predict point-defect concentrations $X_{d}$ in the TaC and HfC
carbides as a function of temperature and chemical composition. For
each defect type $d$, the function $X_{d}(c_{A},T)$ can be conveniently
and accurately represented by two parameters: an Arrhenius prefactor
$X_{d}^{0}$ and effective formation energy $\overline{\varepsilon}_{d}$.
Our results complement the previous work \citep{Smith:2018va,Tang:2020wb,Salehin:2021ti,Razumovskiy:2013uk,Razumovskiy:2015wx}
and can be summarized as follows:
\begin{itemize}
\item Atomic mechanisms of thermal and constitutional disorder in TaC and
HfC are complex and involve multiple types of point defects occurring
simultaneously. 
\item The strong short-range binding among the elementary point defects,
especially in HfC, leads to persistent point-defect clusters. The
dominant type of defect cluster depends on the chemical composition
and temperature and is different between TaC and HfC. 
\item The presence of relatively large concentrations of defect clusters
is one of the hallmarks of these carbides distinguishing them from
other ordered phases such as intermetallic compounds. 
\item The diversity of the point defects suggests that the mechanisms of
metal atom diffusion are complex and may involve chains of point-defect
reactions and collective atomic rearrangements. Their computational
studies require new methods.
\end{itemize}
The mode developed here is general enough to be applied to other binary
B1-ordered compounds and can be further generalized to compounds with
other crystal structures and different stoichiometries.

\bigskip{}

\noindent \textbf{Acknowledgments}

This research was supported by the Office of Naval Research under
Award No. N00014-22-1-2645.

\section*{Appendix: The $L$-numbers and the material balance equation}

The table below presents the complete list of the $L$-numbers for
all defect clusters considered in this work. These numbers appear
in Eq.(\ref{eq:balance}) for the material balance.

\noindent 
\begin{table}[H]
\begin{centering}
\caption{$L$-numbers for the defect clusters considered in this work.}
\bigskip{}
\begin{tabular}{|cr@{\extracolsep{0pt}.}lccccc|}
\hline 
\multirow{1}{*}{Cluster} & \multicolumn{7}{c|}{Elementary defect}\tabularnewline
\hline 
\hline 
 & \multicolumn{2}{c}{$V_{\alpha}$} & $V_{\beta}$  & $A_{\beta}$  & $B_{\alpha}$  & $I_{A}$  & $I_{B}$\tabularnewline
\hline 
$V_{\alpha}V_{\beta}$  & \multicolumn{2}{c}{1} & 1  & 0  & 0  & 0  & 0\tabularnewline
\hline 
$A_{\beta}B_{\alpha}$  & \multicolumn{2}{c}{0} & 0  & 1  & 1  & 0  & 0\tabularnewline
\hline 
$V_{\alpha}B_{\alpha}$  & \multicolumn{2}{c}{1} & 0  & 0  & 1  & 0  & 0\tabularnewline
\hline 
$V_{\beta}A_{\beta}$  & \multicolumn{2}{c}{0} & 1  & 1  & 0  & 0  & 0\tabularnewline
\hline 
$V_{\beta}B_{\alpha}$  & \multicolumn{2}{c}{0} & 1  & 0  & 1  & 0  & 0\tabularnewline
\hline 
$V_{\beta}I_{B}$  & \multicolumn{2}{c}{0} & 1  & 0  & 0  & 0  & 1\tabularnewline
\hline 
$V_{\alpha}V_{\beta}^{2}~(T)$  & \multicolumn{2}{c}{1} & 2  & 0  & 0  & 0  & 0\tabularnewline
\hline 
$V_{\alpha}V_{\beta}^{2}~(\mathrm{L})$  & \multicolumn{2}{c}{1} & 2  & 0  & 0  & 0  & 0\tabularnewline
\hline 
$V_{\alpha}V_{\beta}^{3}~(\mathrm{IP})$  & \multicolumn{2}{c}{1} & 3  & 0  & 0  & 0  & 0\tabularnewline
\hline 
$V_{\alpha}V_{\beta}^{3}~(\mathrm{OP})$  & \multicolumn{2}{c}{1} & 3  & 0  & 0  & 0  & 0\tabularnewline
\hline 
$V_{\alpha}V_{\beta}^{4}~(\mathrm{IP})$  & \multicolumn{2}{c}{1} & 4  & 0  & 0  & 0  & 0\tabularnewline
\hline 
$V_{\alpha}V_{\beta}^{4}~(\mathrm{OP})$  & \multicolumn{2}{c}{1} & 4  & 0  & 0  & 0  & 0\tabularnewline
\hline 
$V_{\alpha}V_{\beta}^{5}$  & \multicolumn{2}{c}{1} & 5  & 0  & 0  & 0  & 0\tabularnewline
\hline 
$V_{\alpha}V_{\beta}^{6}$  & \multicolumn{2}{c}{1} & 6  & 0  & 0  & 0  & 0\tabularnewline
\hline 
\end{tabular}
\par\end{centering}
{\footnotesize{}{}{}\label{table:L-numbers}}{\footnotesize\par}

\end{table}

For the set of defect clusters considered in this work, the material
balance equation is 
\begin{eqnarray}
x & = & \dfrac{1}{4}\left(X_{V_{\beta}}-X_{V_{\alpha}}\right)+\dfrac{1}{2}\left(X_{A_{\beta}}-X_{B_{\alpha}}\right)-\dfrac{1}{2}\nu\left(X_{I_{B}}-X_{I_{A}}\right)\nonumber \\
 & + & \dfrac{1}{4}\left(3X_{V_{\beta}A_{\beta}}-3X_{V_{\alpha}B_{\alpha}}-X_{V_{\beta}B_{\alpha}}\right)\nonumber \\
 & + & \dfrac{1}{4}\left(X_{V_{\alpha}V_{\beta}^{2}(T)}+X_{V_{\alpha}V_{\beta}^{2}(L)}+2X_{V_{\alpha}V_{\beta}^{3}(IP)}+2X_{V_{\alpha}V_{\beta}^{3}(OP)}\right)\nonumber \\
 & + & \dfrac{1}{4}\left(3X_{V_{\alpha}V_{\beta}^{4}(IP)}+3X_{V_{\alpha}V_{\beta}^{4}(OP)}+4X_{V_{\alpha}V_{\beta}^{5}}+5X_{V_{\alpha}V_{\beta}^{6}}\right).\label{eq:balance_1}
\end{eqnarray}
This equation contains 17 defect concentrations. Their calculation
requires solving this equation simultaneously with 16 equations representing
the reactions among the elementary defects and the formation of non-conserving
defect clusters.


\newpage{}

\begin{table}[h]
\noindent \caption{Lattice parameter $a$, cohesive energy $\varepsilon_{0}$, elastic
constants $c_{ij}$, and bulk modulus $B$ obtained by the present
DFT calculations. The experimental and calculated values from the
literature are shown in the round and square brackets, respectively.}
\bigskip{}
\begin{tabular}{|c|c|c|}
\hline 
Property  & TaC  & HfC\tabularnewline
\hline 
\hline 
$a$ (\AA)  & 4.478 {[}4.471{]}$^{a}$, (4.450)$^{b}$  & 4.646 {[}4.642{]}$^{a}$, {[}4.647{]}$^{f}$, (4.631)$^{g}$\tabularnewline
\hline 
$\varepsilon_{0}$ (eV)  & $-11.10$  & $-10.53$\tabularnewline
\hline 
$c_{11}$ (GPa)  & 759.8 {[}737{]}$^{a,c}$, {[}674{]}$^{d}$  & 509.9 {[}577{]}$^{h}$, {[}540{]}$^{a,d}$, (500)$^{i}$\tabularnewline
\hline 
$c_{12}$ (GPa)  & 118.0 {[}141{]}$^{a,c}$, {[}172{]}$^{d}$  & 111.5 {[}117{]}$^{h}$, {[}112{]}$^{a,d}$\tabularnewline
\hline 
$c_{44}$ (GPa)  & 170.8 {[}175{]}$^{a,c}$, {[}167{]}$^{d}$  & 159.2 {[}171{]}$^{h}$, {[}171{]}$^{a,d}$\tabularnewline
\hline 
$B$ (GPa)  & 331.9, (332)$^{e}$, (355.9)$^{b}$, {[}340{]}$^{a,c}$, {[}339{]}$^{d}$  & 244.3, (242)$^{j}$, (272)$^{b}$, {[}253{]}$^{a,d}$\tabularnewline
\hline 
\multicolumn{3}{l}{{\footnotesize{}{}{}$^{a}$Ref.\,\citep{Yu:2015us}, $^{b}$Ref.\,\citep{he2022elasticity},
$^{c}$Ref.\,\citep{Yu:2014uh} , $^{d}$Ref.\,\citep{Smith:2018wa},
$^{e}$Ref.\,\citep{Zhang:2017tx}, $^{f}$Ref.\,\citep{guo2021electronic}}}\tabularnewline
\multicolumn{3}{l}{{\footnotesize{}{}{}$^{g}$Ref.\,\citep{singh1997estimation},
$^{h}$Ref.\,\citep{he2008crystal}, $^{i}$Ref.\,\citep{weber1973lattice},
$^{j}$Ref.\,\citep{brown1966elastic}}}\tabularnewline
\end{tabular}

\noindent \label{table:lattice} 
\end{table}

\begin{table}[h]
\caption{``Raw'' energies (in eV) of point defects in TaC and HfC obtained
by DFT calculations. The defect pairs are considered to have the shortest
defect separation. L, T, IP, and OP are vacancy cluster configurations
explained in the main text. {*}Unstable configuration.}
\bigskip{}
\begin{tabular}{|l|c|r@{\extracolsep{0pt}.}l|r@{\extracolsep{0pt}.}l|}
\hline 
Defect type  & Symbol  & \multicolumn{2}{c|}{TaC} & \multicolumn{2}{c|}{HfC}\tabularnewline
\hline 
\hline 
Metal vacancy  & $V_{\alpha}$  & \multicolumn{2}{c|}{$15.30$} & \multicolumn{2}{c|}{$19.30$}\tabularnewline
\hline 
Carbon vacancy  & $V_{\beta}$  & \multicolumn{2}{c|}{$9.45$} & \multicolumn{2}{c|}{$10.20$}\tabularnewline
\hline 
Antisite on carbon sublattice  & $A_{\beta}$  & \multicolumn{2}{c|}{$7.59$} & \multicolumn{2}{c|}{$9.13$}\tabularnewline
\hline 
Antisite on metal sublattice  & $B_{\alpha}$  & \multicolumn{2}{c|}{$11.78$} & \multicolumn{2}{c|}{$13.80$}\tabularnewline
\hline 
Metal interstitial  & $I_{A}$  & \multicolumn{2}{c|}{$-0.27$} & \multicolumn{2}{c|}{$-0.31$}\tabularnewline
\hline 
Carbon interstitial  & $I_{B}$  & \multicolumn{2}{c|}{$-2.71$} & \multicolumn{2}{c|}{$-3.99$}\tabularnewline
\hline 
Divacancy  & $V_{\alpha}V_{\beta}$  & \multicolumn{2}{c|}{$24.61$} & \multicolumn{2}{c|}{$27.90$}\tabularnewline
\hline 
Antisite pair  & $A_{\beta}B_{\alpha}$  & \multicolumn{2}{c|}{$5.45$} & \multicolumn{2}{c|}{$5.00$}\tabularnewline
\hline 
Antisite-vacancy pair on metal sublattice  & $V_{\alpha}B_{\alpha}$  & \multicolumn{2}{c|}{$22.63$} & \multicolumn{2}{c|}{$25.86$}\tabularnewline
\hline 
Antisite-vacancy pair on carbon sublattice  & $V_{\beta}A_{\beta}$  & \multicolumn{2}{c|}{$15.62$} & \multicolumn{2}{c|}{$18.27$}\tabularnewline
\hline 
Antisite-vacancy pair on different sublattices  & $V_{\beta}B_{\alpha}$  & \multicolumn{2}{c|}{{*}} & \multicolumn{2}{c|}{$20.46$}\tabularnewline
\hline 
Nearest-neighbor Frenkel defect  & $V_{\beta}I_{B}$  & \multicolumn{2}{c|}{{*}} & \multicolumn{2}{c|}{$4.30$}\tabularnewline
\hline 
Vacancy cluster (T)  & $V_{\alpha}V_{\beta}^{2}$  & \multicolumn{2}{c|}{$34.08$} & \multicolumn{2}{c|}{$36.80$}\tabularnewline
\hline 
Vacancy cluster (L)  & $V_{\alpha}V_{\beta}^{2}$  & \multicolumn{2}{c|}{$33.94$} & \multicolumn{2}{c|}{$37.03$}\tabularnewline
\hline 
Vacancy cluster (IP)  & $V_{\alpha}V_{\beta}^{3}$  & \multicolumn{2}{c|}{$43.60$} & \multicolumn{2}{c|}{$46.01$}\tabularnewline
\hline 
Vacancy cluster (OP)  & $V_{\alpha}V_{\beta}^{3}$  & \multicolumn{2}{c|}{$43.62$} & \multicolumn{2}{c|}{$46.17$}\tabularnewline
\hline 
Vacancy cluster (IP)  & $V_{\alpha}V_{\beta}^{4}$  & \multicolumn{2}{c|}{$53.16$} & \multicolumn{2}{c|}{$55.39$}\tabularnewline
\hline 
Vacancy cluster (OP)  & $V_{\alpha}V_{\beta}^{4}$  & \multicolumn{2}{c|}{$53.13$} & \multicolumn{2}{c|}{$55.75$}\tabularnewline
\hline 
Vacancy cluster  & $V_{\alpha}V_{\beta}^{5}$  & \multicolumn{2}{c|}{$62.6$} & \multicolumn{2}{c|}{$65.19$}\tabularnewline
\hline 
Vacancy cluster  & $V_{\alpha}V_{\beta}^{6}$  & \multicolumn{2}{c|}{$72.13$} & \multicolumn{2}{c|}{$75.01$}\tabularnewline
\hline 
\end{tabular}\label{table:raw} 
\end{table}

\begin{table}[h]
\caption{Binding energies and symmetry factors $\sigma$ of defect clusters
in TaC and HfC obtained by DFT calculations. L, T, IP, and OP are
vacancy cluster configurations explained in the main text. {*}Unstable
configuration. The values in square brackets refer to previous calculations.}
\bigskip{}
\begin{tabular}{|c|c|r@{\extracolsep{0pt}.}l|r@{\extracolsep{0pt}.}lr@{\extracolsep{0pt}.}l|c|}
\hline 
\multirow{1}{*}{Cluster type} & \multicolumn{3}{c|}{TaC} & \multicolumn{5}{c|}{HfC}\tabularnewline
\hline 
\hline 
 & $\sigma$  & \multicolumn{2}{c|}{Energy (eV)} & \multicolumn{2}{c}{$\sigma$} & \multicolumn{2}{c|}{} & Energy (eV)\tabularnewline
\hline 
$V_{\alpha}V_{\beta}$  & 6  & \multicolumn{2}{c|}{$-0.14$ {[}$-0.16${]}$^{a}$} & \multicolumn{2}{c}{6} & \multicolumn{2}{c|}{} & $-1.60$ {[}$-1.54${]}$^{a}$, {[}$-1.49${]}$^{b}$,\tabularnewline
\hline 
$A_{\beta}B_{\alpha}$  & 6  & \multicolumn{2}{c|}{$-13.92$} & \multicolumn{2}{c}{6} & \multicolumn{2}{c|}{} & $-17.93$\tabularnewline
\hline 
$V_{\alpha}B_{\alpha}$  & 6  & \multicolumn{2}{c|}{$-4.45$} & \multicolumn{2}{c}{6} & \multicolumn{2}{c|}{} & $-7.24$\tabularnewline
\hline 
$V_{\beta}A_{\beta}$  & 6  & \multicolumn{2}{c|}{$-1.42$} & \multicolumn{2}{c}{12} & \multicolumn{2}{c|}{} & $-1.06$\tabularnewline
\hline 
$V_{\beta}B_{\alpha}$  &  & \multicolumn{2}{c|}{{*}} & \multicolumn{2}{c}{6} & \multicolumn{2}{c|}{} & $-3.54$\tabularnewline
\hline 
$V_{\beta}I_{B}$  &  & \multicolumn{2}{c|}{{*}} & \multicolumn{2}{c}{8} & \multicolumn{2}{c|}{} & $-1.91$\tabularnewline
\hline 
$V_{\alpha}V_{\beta}^{2}~(T)$  & 12  & \multicolumn{2}{c|}{$-0.12$} & \multicolumn{2}{c}{12} & \multicolumn{2}{c|}{} & $-2.90$\tabularnewline
\hline 
$V_{\alpha}V_{\beta}^{2}~(\mathrm{L})$  & 3  & \multicolumn{2}{c|}{$-0.26$} & \multicolumn{2}{c}{3} & \multicolumn{2}{c|}{} & $-2.67$\tabularnewline
\hline 
$V_{\alpha}V_{\beta}^{3}~(\mathrm{IP})$  & 12  & \multicolumn{2}{c|}{$-0.05$} & \multicolumn{2}{c}{12} & \multicolumn{2}{c|}{} & $-3.89$\tabularnewline
\hline 
$V_{\alpha}V_{\beta}^{3}~(\mathrm{OP})$  & 8  & \multicolumn{2}{c|}{$-0.03$} & \multicolumn{2}{c}{8} & \multicolumn{2}{c|}{} & $-3.73$\tabularnewline
\hline 
$V_{\alpha}V_{\beta}^{4}~(\mathrm{IP})$  & 3  & \multicolumn{2}{c|}{$0.06$} & \multicolumn{2}{c}{3} & \multicolumn{2}{c|}{} & $-4.71$\tabularnewline
\hline 
$V_{\alpha}V_{\beta}^{4}~(\mathrm{OP})$  & 12  & \multicolumn{2}{c|}{$0.03$} & \multicolumn{2}{c}{12} & \multicolumn{2}{c|}{} & $-4.35$\tabularnewline
\hline 
$V_{\alpha}V_{\beta}^{5}$  & 6  & \multicolumn{2}{c|}{$0.05$} & \multicolumn{2}{c}{6} & \multicolumn{2}{c|}{} & $-5.11$\tabularnewline
\hline 
$V_{\alpha}V_{\beta}^{6}$  & 1  & \multicolumn{2}{c|}{$0.13$} & \multicolumn{2}{c}{1} & \multicolumn{2}{c|}{} & $-5.49$ {[}$-5.52${]}$^{b}$\tabularnewline
\hline 
\multicolumn{2}{l}{$^{a}$Ref.~\citep{Salehin:2021ti}, $^{b}$Ref.~\citep{Razumovskiy:2015wx}} & \multicolumn{2}{c}{} & \multicolumn{2}{c}{} & \multicolumn{3}{c}{}\tabularnewline
\end{tabular}

{\footnotesize{}{}{}\label{table:bindingenergy}}{\footnotesize\par}
\end{table}

\begin{table}
\caption{Equations for the Arrhenius parameters (prefactors $X_{d}$ and effective
formation energies $\overline{\varepsilon}_{d}$) of point defects
in metal-rich carbides ($x>0$) with constitutional defects $V_{\beta}$.\label{tab:Arrhenius_eqns_x}}
\bigskip{}
\centering{}%
\begin{tabular}{|c|c|c|}
\hline 
Defect & $X_{d}^{0}$ & $\overline{\varepsilon}_{d}$\tabularnewline
\hline 
\hline 
$V_{\alpha}$ & $\dfrac{1}{4x}$ & $\varepsilon_{V_{\alpha}}+\varepsilon_{V_{\beta}}+2\varepsilon_{0}$\tabularnewline
\hline 
$V_{\beta}$ & $4x$ & $0$\tabularnewline
\hline 
$A_{\beta}$ & $(4x)^{2}$ & $\varepsilon_{A_{\beta}}-2\varepsilon_{V_{\beta}}-2\varepsilon_{0}$\tabularnewline
\hline 
$B_{\alpha}$ & $\dfrac{1}{(4x)^{2}}$ & $\varepsilon_{B_{\alpha}}+2\varepsilon_{V_{\beta}}+2\varepsilon_{0}$\tabularnewline
\hline 
$I_{A}$ & $4x$ & $\varepsilon_{I_{A}}-\varepsilon_{V_{\beta}}-2\varepsilon_{0}$\tabularnewline
\hline 
$I_{B}$ & $\dfrac{1}{4x}$ & $\varepsilon_{I_{B}}+\varepsilon_{V_{\beta}}$\tabularnewline
\hline 
$A_{\beta}B_{\alpha}$ & $\sigma_{A_{\beta}B_{\alpha}}$ & $\varepsilon_{A_{\beta}B_{\alpha}}$\tabularnewline
\hline 
$V_{\alpha}B_{\alpha}$ & $\dfrac{\sigma_{V_{\alpha}B_{\alpha}}}{(4x)^{3}}$ & $\varepsilon_{V_{\alpha}B_{\alpha}}+3\varepsilon_{V_{\beta}}+4\varepsilon_{0}$\tabularnewline
\hline 
$V_{\beta}A_{\beta}$ & $\sigma_{V_{\beta}B_{\beta}}(4x)^{3}$ & $\varepsilon_{V_{\beta}B_{\beta}}-3\varepsilon_{V_{\beta}}-2\varepsilon_{0}$\tabularnewline
\hline 
$V_{\beta}B_{\alpha}$ & $\dfrac{\sigma_{V_{\beta}B_{\alpha}}}{4x}$ & $\varepsilon_{V_{\beta}B_{\alpha}}+\varepsilon_{V_{\beta}}+2\varepsilon_{0}$\tabularnewline
\hline 
$V_{\beta}I_{B}$ & $\sigma_{V_{\beta}I_{B}}$ & $\varepsilon_{V_{\beta}I_{B}}$\tabularnewline
\hline 
$V_{\alpha}V_{\beta}$ & $\sigma_{V_{\alpha}V_{\beta}}$ & $\varepsilon_{V_{\alpha}V_{\beta}}+2\varepsilon_{0}$\tabularnewline
\hline 
$V_{\alpha}V_{\beta}^{2}$ & $\sigma_{V_{\alpha}V_{\beta}^{2}}(4x)$ & $\varepsilon_{V_{\alpha}V_{\beta}^{2}}-\varepsilon_{V_{\beta}}+2\varepsilon_{0}$\tabularnewline
\hline 
$V_{\alpha}V_{\beta}^{3}$ & $\sigma_{V_{\alpha}V_{\beta}^{3}}(4x)^{2}$ & $\varepsilon_{V_{\alpha}V_{\beta}^{3}}-2\varepsilon_{V_{\beta}}+2\varepsilon_{0}$\tabularnewline
\hline 
$V_{\alpha}V_{\beta}^{4}$ & $\sigma_{V_{\alpha}V_{\beta}^{4}}(4x)^{3}$ & $\varepsilon_{V_{\alpha}V_{\beta}^{4}}-3\varepsilon_{V_{\beta}}+2\varepsilon_{0}$\tabularnewline
\hline 
$V_{\alpha}V_{\beta}^{5}$ & $\sigma_{V_{\alpha}V_{\beta}^{5}}(4x)^{4}$ & $\varepsilon_{V_{\alpha}V_{\beta}^{5}}-4\varepsilon_{V_{\beta}}+2\varepsilon_{0}$\tabularnewline
\hline 
$V_{\alpha}V_{\beta}^{6}$ & $\sigma_{V_{\alpha}V_{\beta}^{6}}(4x)^{5}$ & $\varepsilon_{V_{\alpha}V_{\beta}^{6}}-5\varepsilon_{V_{\beta}}+2\varepsilon_{0}$\tabularnewline
\hline 
\end{tabular}
\end{table}

\begin{table}
\centering{}\caption{Equations for the Arrhenius parameters (prefactors $X_{d}$ and effective
formation energies $\overline{\varepsilon}_{d}$) of point defects
in stoichiometric carbides ($x=0$). Two mechanisms of thermal disorder
are considered with the leading defects $V_{\alpha}$ \& $V_{\beta}$
and $V_{\beta}$ \& $I_{B}$. \label{tab:Arrhenius_eqns_x0}}
\bigskip{}
\begin{tabular}{|c|c|c|c|c|}
\hline 
 & \multicolumn{2}{c|}{Leading defects $V_{\alpha}$ \& $V_{\beta}$} & \multicolumn{2}{c|}{Leading defects $V_{\beta}$ \& $I_{B}$}\tabularnewline
\hline 
\hline 
Defect & $X_{d}^{0}$ & $\overline{\varepsilon}_{d}$ & $X_{d}^{0}$ & $\overline{\varepsilon}_{d}$\tabularnewline
\hline 
\hline 
$V_{\alpha}$ & 1 & $\dfrac{\varepsilon_{V_{\alpha}}+\varepsilon_{V_{\beta}}+2\varepsilon_{0}}{2}$ & $\dfrac{1}{\sqrt{2\nu}}$ & $\varepsilon_{V_{\alpha}}+2\varepsilon_{0}+\dfrac{\varepsilon_{V_{\beta}}-\varepsilon_{I_{B}}}{2}$\tabularnewline
\hline 
$V_{\beta}$ & 1 & $\dfrac{\varepsilon_{V_{\alpha}}+\varepsilon_{V_{\beta}}+2\varepsilon_{0}}{2}$ & $\sqrt{2\nu}$ & $\dfrac{\varepsilon_{V_{\beta}}+\varepsilon_{I_{B}}}{2}$\tabularnewline
\hline 
$A_{\beta}$ & 1 & $\varepsilon_{A_{\beta}}-\varepsilon_{V_{\beta}}+\varepsilon_{V_{\alpha}}$ & $2\nu$ & $\varepsilon_{A_{\beta}}-\varepsilon_{V_{\beta}}+\varepsilon_{I_{B}}-2\varepsilon_{0}$\tabularnewline
\hline 
$B_{\alpha}$ & 1 & $\varepsilon_{B_{\alpha}}+\varepsilon_{V_{\beta}}-\varepsilon_{V_{\alpha}}$ & $\dfrac{1}{2\nu}$ & $\varepsilon_{B_{\alpha}}+\varepsilon_{V_{\beta}}-\varepsilon_{I_{B}}+2\varepsilon_{0}$\tabularnewline
\hline 
$I_{A}$ & 1 & $\varepsilon_{I_{A}}-\varepsilon_{0}+\dfrac{\varepsilon_{V_{\alpha}}-\varepsilon_{V_{\beta}}}{2}$ & $\sqrt{2\nu}$ & $\varepsilon_{I_{A}}-2\varepsilon_{0}+\dfrac{\varepsilon_{I_{B}}-\varepsilon_{V_{\beta}}}{2}$\tabularnewline
\hline 
$I_{B}$ & 1 & $\varepsilon_{I_{B}}-\varepsilon_{0}+\dfrac{\varepsilon_{V_{\beta}}-\varepsilon_{V_{\alpha}}}{2}$ & $\dfrac{1}{\sqrt{2\nu}}$ & $\dfrac{\varepsilon_{V_{\beta}}+\varepsilon_{I_{B}}}{2}$\tabularnewline
\hline 
$A_{\beta}B_{\alpha}$ & $\sigma_{A_{\beta}B_{\alpha}}$ & $\varepsilon_{A_{\beta}B_{\alpha}}$ & $\sigma_{A_{\beta}B_{\alpha}}$ & $\varepsilon_{A_{\beta}B_{\alpha}}$\tabularnewline
\hline 
$V_{\alpha}B_{\alpha}$ & $\sigma_{V_{\alpha}B_{\alpha}}$ & $\varepsilon_{V_{\alpha}B_{\alpha}}+\varepsilon_{0}+\dfrac{3(\varepsilon_{V_{\beta}}-\varepsilon_{V_{\alpha})}}{2}$ & $\dfrac{\sigma_{V_{\alpha}B_{\alpha}}}{(2\nu)^{3/2}}$ & $\varepsilon_{V_{\alpha}B_{\alpha}}+4\varepsilon_{0}+\dfrac{3(\varepsilon_{V_{\beta}}-\varepsilon_{I_{B})}}{2}$\tabularnewline
\hline 
$V_{\beta}A_{\beta}$ & $\sigma_{V_{\beta}B_{\beta}}$ & $\varepsilon_{V_{\beta}B_{\beta}}+\varepsilon_{0}+\dfrac{3(\varepsilon_{V_{\alpha}}-\varepsilon_{V_{\beta}})}{2}$ & $(2\nu)^{3/2}\sigma_{V_{\beta}B_{\beta}}$ & $\varepsilon_{V_{\beta}A_{\beta}}-2\varepsilon_{0}-\dfrac{3(\varepsilon_{V_{\beta}}-\varepsilon_{I_{B})}}{2}$\tabularnewline
\hline 
$V_{\beta}B_{\alpha}$ & $\sigma_{V_{\beta}B_{\alpha}}$ & $\varepsilon_{V_{\beta}B_{\alpha}}+\varepsilon_{0}+\dfrac{\varepsilon_{V_{\beta}}-\varepsilon_{V_{\alpha}}}{2}$ & $\dfrac{\sigma_{V_{\beta}B_{\alpha}}}{\sqrt{2\nu}}$ & $\varepsilon_{V_{\beta}B_{\alpha}}+2\varepsilon_{0}+\dfrac{\varepsilon_{V_{\beta}}-\varepsilon_{I_{B}}}{2}$\tabularnewline
\hline 
$V_{\beta}I_{B}$ & $\sigma_{V_{\beta}I_{B}}$ & $\varepsilon_{V_{\beta}I_{B}}$ & $\sigma_{V_{\beta}I_{B}}$ & $\varepsilon_{V_{\beta}I_{B}}$\tabularnewline
\hline 
$V_{\alpha}V_{\beta}$ & $\sigma_{V_{\alpha}V_{\beta}}$ & $\varepsilon_{V_{\alpha}V_{\beta}}+2\varepsilon_{0}$ & $\sigma_{V_{\alpha}V_{\beta}}$ & $\varepsilon_{V_{\alpha}V_{\beta}}+2\varepsilon_{0}$\tabularnewline
\hline 
$V_{\alpha}V_{\beta}^{2}$ & $\sigma_{V_{\alpha}V_{\beta}^{2}}$ & $\varepsilon_{V_{\alpha}V_{\beta}^{2}}+3\varepsilon_{0}+\dfrac{\varepsilon_{V_{\alpha}}-\varepsilon_{V_{\beta}}}{2}$ & $\sqrt{2\nu}\sigma_{V_{\alpha}V_{\beta}^{2}}$ & $\varepsilon_{V_{\alpha}V_{\beta}^{2}}+2\varepsilon_{0}+\dfrac{\varepsilon_{I_{B}}-\varepsilon_{V_{\beta}}}{2}$\tabularnewline
\hline 
$V_{\alpha}V_{\beta}^{3}$ & $\sigma_{V_{\alpha}V_{\beta}^{3}}$ & $\varepsilon_{V_{\alpha}V_{\beta}^{3}}+(\varepsilon_{V_{\alpha}}-\varepsilon_{V_{\beta}})+4\varepsilon_{0}$ & $2\nu\sigma_{V_{\alpha}V_{\beta}^{3}}$ & $\varepsilon_{V_{\alpha}V_{\beta}^{3}}+2\varepsilon_{0}+\varepsilon_{I_{B}}-\varepsilon_{V_{\beta}}$\tabularnewline
\hline 
$V_{\alpha}V_{\beta}^{4}$ & $\sigma_{V_{\alpha}V_{\beta}^{4}}$ & $\varepsilon_{V_{\alpha}V_{\beta}^{4}}+5\varepsilon_{0}+\dfrac{3(\varepsilon_{V_{\alpha}}-\varepsilon_{V_{\beta}})}{2}$ & $(2\nu)^{3/2}\sigma_{V_{\alpha}V_{\beta}^{4}}$ & $\varepsilon_{V_{\alpha}V_{\beta}^{4}}+2\varepsilon_{0}+\dfrac{3(\varepsilon_{I_{B}}-\varepsilon_{V_{\beta}})}{2}$\tabularnewline
\hline 
$V_{\alpha}V_{\beta}^{5}$ & $\sigma_{V_{\alpha}V_{\beta}^{5}}$ & $\varepsilon_{V_{\alpha}V_{\beta}^{5}}+2(\varepsilon_{V_{\alpha}}-\varepsilon_{V_{\beta}})+6\varepsilon_{0}$ & $(2\nu)^{2}\sigma_{V_{\alpha}V_{\beta}^{5}}$ & $\varepsilon_{V_{\alpha}V_{\beta}^{5}}+2\varepsilon_{0}+2(\varepsilon_{I_{B}}-\varepsilon_{V_{\beta}})$\tabularnewline
\hline 
$V_{\alpha}V_{\beta}^{6}$ & $\sigma_{V_{\alpha}V_{\beta}^{6}}$ & $\varepsilon_{V_{\alpha}V_{\beta}^{6}}+7\varepsilon_{0}+\dfrac{5(\varepsilon_{V_{\alpha}}-\varepsilon_{V_{\beta}})}{2}$ & $(2\nu)^{5/2}\sigma_{V_{\alpha}V_{\beta}^{6}}$ & $\varepsilon_{V_{\alpha}V_{\beta}^{6}}+2\varepsilon_{0}+\dfrac{5(\varepsilon_{I_{B}}-\varepsilon_{V_{\beta}})}{2}$\tabularnewline
\hline 
\end{tabular}
\end{table}

\begin{table}
\caption{Arrhenius parameters (prefactors $X_{d}$ and effective formation
energies $\overline{\varepsilon}_{d}$) of point defects in TaC and
HfC with stoichiometric ($x=0$) and metal-rich ($x>0$) compositions.
The values in square brackets were computed from the literature reports.
\label{tab:Arrhenius-parameters}}
\bigskip{}
\begin{tabular}{|c|c|c|c|c|c|c|c|c|}
\hline 
 & \multicolumn{4}{c|}{TaC} & \multicolumn{4}{c|}{HfC}\tabularnewline
\hline 
\hline 
 & \multicolumn{2}{c|}{$x=0$, $V_{C}$ \& $V_{Ta}$} & \multicolumn{2}{c|}{$x>0$, $V_{C}$} & \multicolumn{2}{c|}{$x=0$, $V_{C}$ \& $I_{C}$} & \multicolumn{2}{c|}{$x>0$, $V_{C}$}\tabularnewline
\hline 
Defect & $X_{d}^{0}$ & $\overline{\varepsilon}_{d}$ (eV) & $X_{d}^{0}$ & $\overline{\varepsilon}_{d}$ (eV) & $X_{d}^{0}$ & $\overline{\varepsilon}_{d}$ (eV) & $X_{d}^{0}$ & $\overline{\varepsilon}_{d}$ (eV)\tabularnewline
\hline 
$V_{\alpha}$ & 1 & 1.28 & $\dfrac{1}{4x}$ & 2.55 {[}$2.65${]}$^{b}$ & $\dfrac{1}{\sqrt{2\nu}}$ & 5.34 & $\dfrac{1}{4x}$ & 8.44 {[}$8.57${]}$^{a}$\tabularnewline
 &  &  &  &  &  &  &  & {[}$8.64${]}$^{b}$\tabularnewline
\hline 
$V_{\beta}$ & 1 & 1.28 & $4x$ & 0.00 & $\sqrt{2\nu}$ & 3.11 & $4x$ & 0.00\tabularnewline
\hline 
$A_{\beta}$ & 1 & 13.44 & $16x^{2}$ & 10.90 & $2\nu$ & 16.00 & $16x^{2}$ & 9.79\tabularnewline
\hline 
$B_{\alpha}$ & 1 & 5.93 & $\dfrac{1}{16x^{2}}$ & 8.48 & $\dfrac{1}{2\nu}$ & 6.93 & $\dfrac{1}{16x^{2}}$ & 13.14 {[}$13.51${]}$^{a}$\tabularnewline
\hline 
$I_{A}$ & 1 & 13.76 & $4x$ & 12.48 & $\sqrt{2\nu}$ & 13.66 & $4x$ & 10.55\tabularnewline
\hline 
$I_{B}$ & 1 & 5.47 & $\dfrac{1}{4x}$ & 6.74 & $\dfrac{1}{\sqrt{2\nu}}$ & 3.11 & $\dfrac{1}{4x}$ & 6.21\tabularnewline
\hline 
$A_{\beta}B_{\alpha}$ & 6 & 5.45 & 6 & 5.45 & 6 & 5.00 & 6 & 5.00\tabularnewline
\hline 
$V_{\alpha}B_{\alpha}$ & 6 & 2.76 & $\dfrac{6}{64x^{3}}$ & 6.58 & $\dfrac{6}{(2\nu)^{3/2}}$ & 5.03 & $\dfrac{6}{64x^{3}}$ & 14.34\tabularnewline
\hline 
$V_{\beta}A_{\beta}$ & 6 & 13.30 & $384x^{3}$ & 9.47 & $12(2\nu)^{3/2}$ & 18.05 & $384x^{3}$ & 8.73\tabularnewline
\hline 
$V_{\beta}B_{\alpha}$ & {*} & {*} & {*} &  & $\dfrac{6}{\sqrt{2\nu}}$ & 6.50 & $\dfrac{6}{4x}$ & 9.60\tabularnewline
\hline 
$V_{\beta}I_{B}$ & {*} & {*} & {*} &  & 8 & 4.30 & 8 & 4.30\tabularnewline
\hline 
$V_{\alpha}V_{\beta}$ & 6 & 2.41 & 6 & 2.41 {[}$2.49${]}$^{b}$ & 6 & 6.84 & 6 & 6.84 {[}$7.08${]}$^{a}$\tabularnewline
\hline 
$V_{\alpha}V_{\beta}^{2}$ (T) & 12 & 3.71 & $48x$ & 2.43 {[}$2.50${]}$^{b}$ & $12\sqrt{2\nu}$ & 8.65 & $48x$ & 5.54 {[}$5.85${]}$^{b}$\tabularnewline
\hline 
$V_{\alpha}V_{\beta}^{2}$ (L) & 3 & 3.57 & $12x$ & 2.29 {[}$2.36${]}$^{b}$ & $3\sqrt{2\nu}$ & 8.88 & $12x$ & 5.77 {[}$6.05${]}$^{b}$\tabularnewline
\hline 
$V_{\alpha}V_{\beta}^{3}$ (IP) & 12 & 5.05 & $192x^{2}$ & 2.50 & $24\nu$ & 10.76 & $192x^{2}$ & 4.55 {[}$4.31${]}$^{b}$\tabularnewline
\hline 
$V_{\alpha}V_{\beta}^{3}$ (OP) & 8 & 5.07 & $128x^{2}$ & 2.52 {[}$2.55${]}$^{b}$ & $16\nu$ & 10.92 & $128x^{2}$ & 4.71 {[}$5.00${]}$^{b}$\tabularnewline
\hline 
$V_{\alpha}V_{\beta}^{4}$ (IP) & 3 & 6.44 & $192x^{3}$ & 2.61 {[}$2.65${]}$^{b}$ & $3(2\nu)^{3/2}$ & 13.05 & $192x^{3}$ & 3.73\tabularnewline
\hline 
$V_{\alpha}V_{\beta}^{4}$ (OP) & 12 & 6.41 & $768x^{3}$ & 2.58 & $12(2\nu)^{3/2}$ & 13.41 & $768x^{3}$ & 4.10 {[}$4.18${]}$^{b}$,\tabularnewline
\hline 
$V_{\alpha}V_{\beta}^{5}$ & 6 & 7.70 & $1536x^{4}$ & 2.60 {[}$2.70${]}$^{b}$ & $6(2\nu)^{2}$ & 15.75 & $1536x^{4}$ & 3.33 {[}$3.45${]}$^{b}$,\tabularnewline
\hline 
$V_{\alpha}V_{\beta}^{6}$ & 1 & 9.06 & $1024x^{5}$ & 2.68 {[}$2.78${]}$^{b}$ & $(2\nu)^{5/2}$ & 18.47 & $1024x^{5}$ & 2.95 {[}$2.95${]}$^{b}$\tabularnewline
 &  &  &  &  &  &  &  & {[}$3.05${]}$^{a}$\tabularnewline
\hline 
\multicolumn{4}{l}{$^{a}$Ref. \citep{Razumovskiy:2015wx} , $^{b}$Ref.~~\citep{Salehin:2021ti}} & \multicolumn{3}{c}{} & \multicolumn{1}{c}{} & \multicolumn{1}{c}{}\tabularnewline
\end{tabular}
\end{table}

\newpage{}

\clearpage\newpage{}
\begin{center}
\setcounter{figure}{0}\setcounter{equation}{0}\setcounter{page}{1}
\par\end{center}

\begin{center}
\textbf{\large{}SUPPLEMENTARY INFORMATION}{\large\par}
\par\end{center}

\bigskip{}

\textbf{\large{}First-principles prediction of point defect energies
and concentrations in the tantalum and hafnium carbides}{\large\par}

\bigskip{}

\begin{quote}
\begin{center}
{\large{}I. Khatri, R. K. Koju, and Y. Mishin}{\large\par}
\par\end{center}

\end{quote}
{\large{}\bigskip{}
}{\large\par}
\begin{center}
{\large{}Department of Physics and Astronomy, MSN 3F3, George Mason
University, Fairfax, Virginia 22030, USA}{\large\par}
\par\end{center}

\begin{figure}[H]
\begin{centering}
\includegraphics[width=0.8\textwidth]{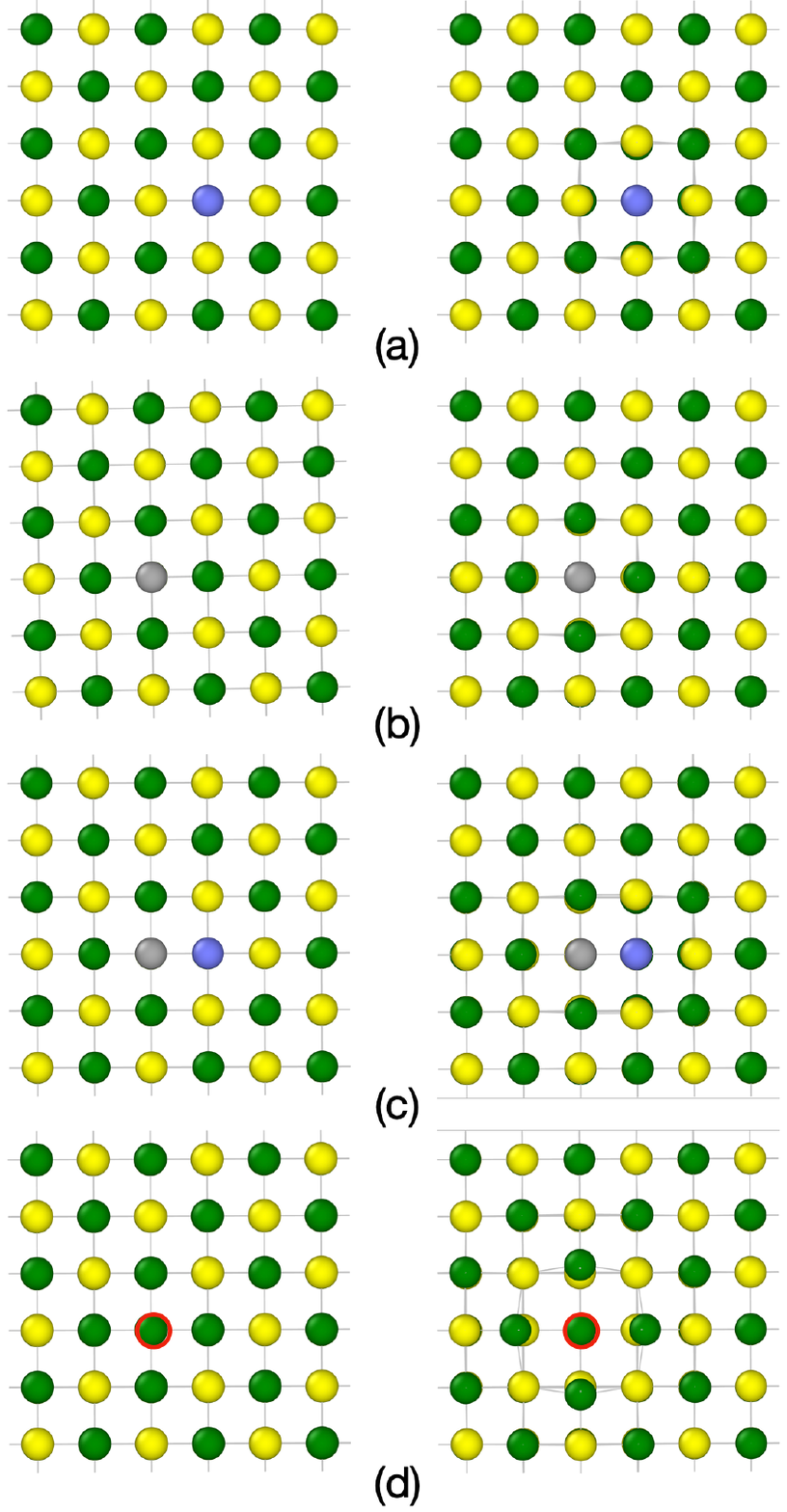}
\par\end{centering}
\caption{Initial (left column) and relaxed (right column) structures of point-defects
undergoing structural reconstructions. (a) $V_{\alpha}$ (b) $V_{\beta}$,
(c) $V_{\alpha}V_{\beta}$, and (d) $A_{\beta}$ in TaC and HfC. The
structures are viewed along a $\left\langle 100\right\rangle $ direction.
The metal atoms, carbon atoms, metal vacancy and carbon vacancy are
shown in green, yellow, purple, and light gray, respectively. Some
atoms are encircled in red for tracking. \label{fig:relaxed_structures-Iso}}
\
\end{figure}

\begin{figure}
\begin{centering}
\includegraphics[width=0.8\textwidth]{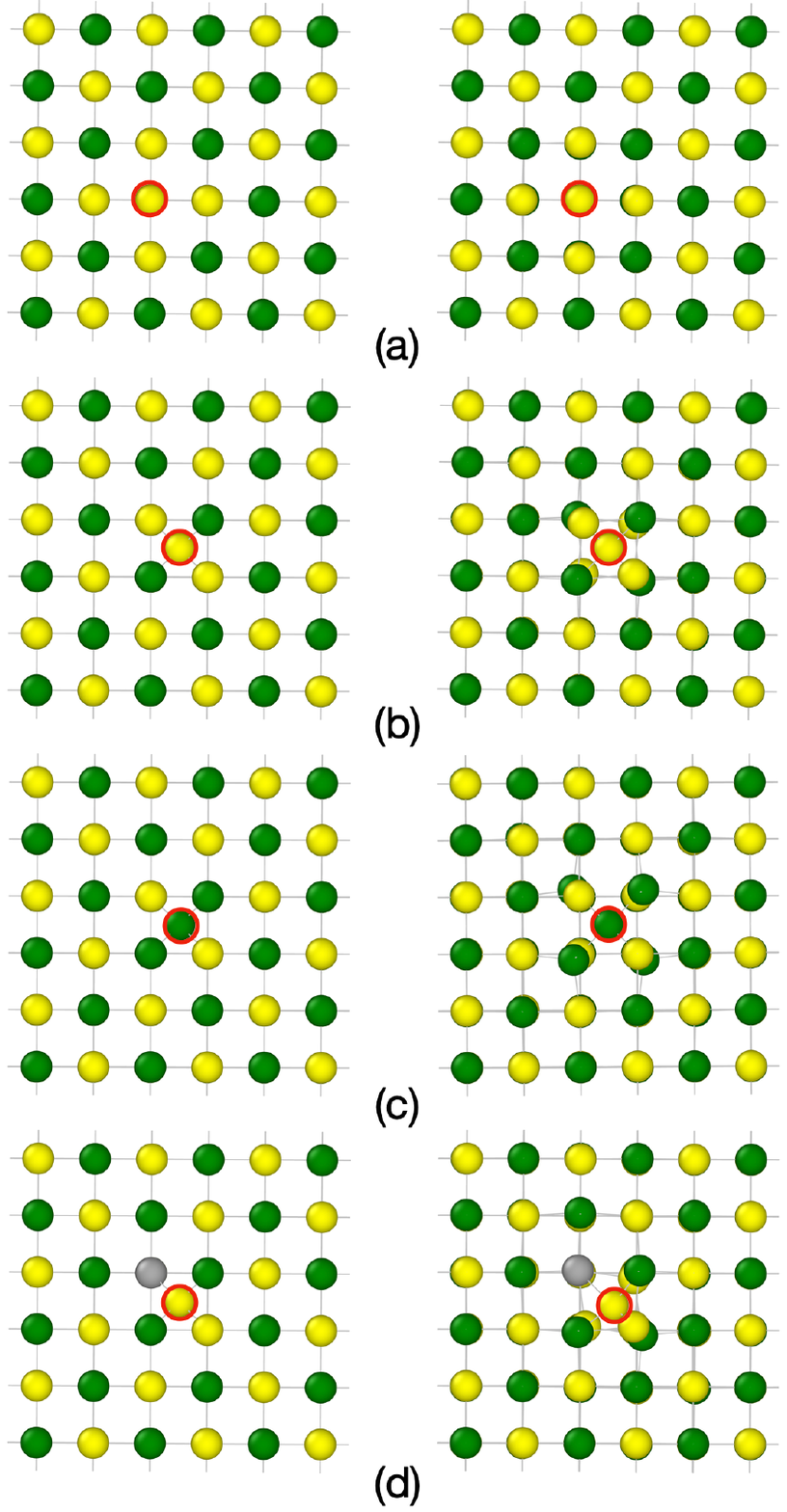}
\par\end{centering}
\caption{Initial (left column) and relaxed (right column) structures of point-defects
undergoing structural reconstructions. (a) $B_{\alpha}$ in TaC and
HfC, (b) $I_{B}$ in TaC and HfC, (c) $I_{A}$ in TaC and HfC, and
(d) $V_{\beta}I_{B}$ in HfC. The structures are viewed along a $\left\langle 100\right\rangle $
direction. The metal atoms, carbon atoms, metal vacancy and carbon
vacancy are shown in green, yellow, purple, and light gray, respectively.
Some atoms are encircled in red for tracking. \label{fig:relaxed_structures-Iso}}
\
\end{figure}

\begin{figure}
\begin{centering}
\includegraphics[width=0.8\textwidth]{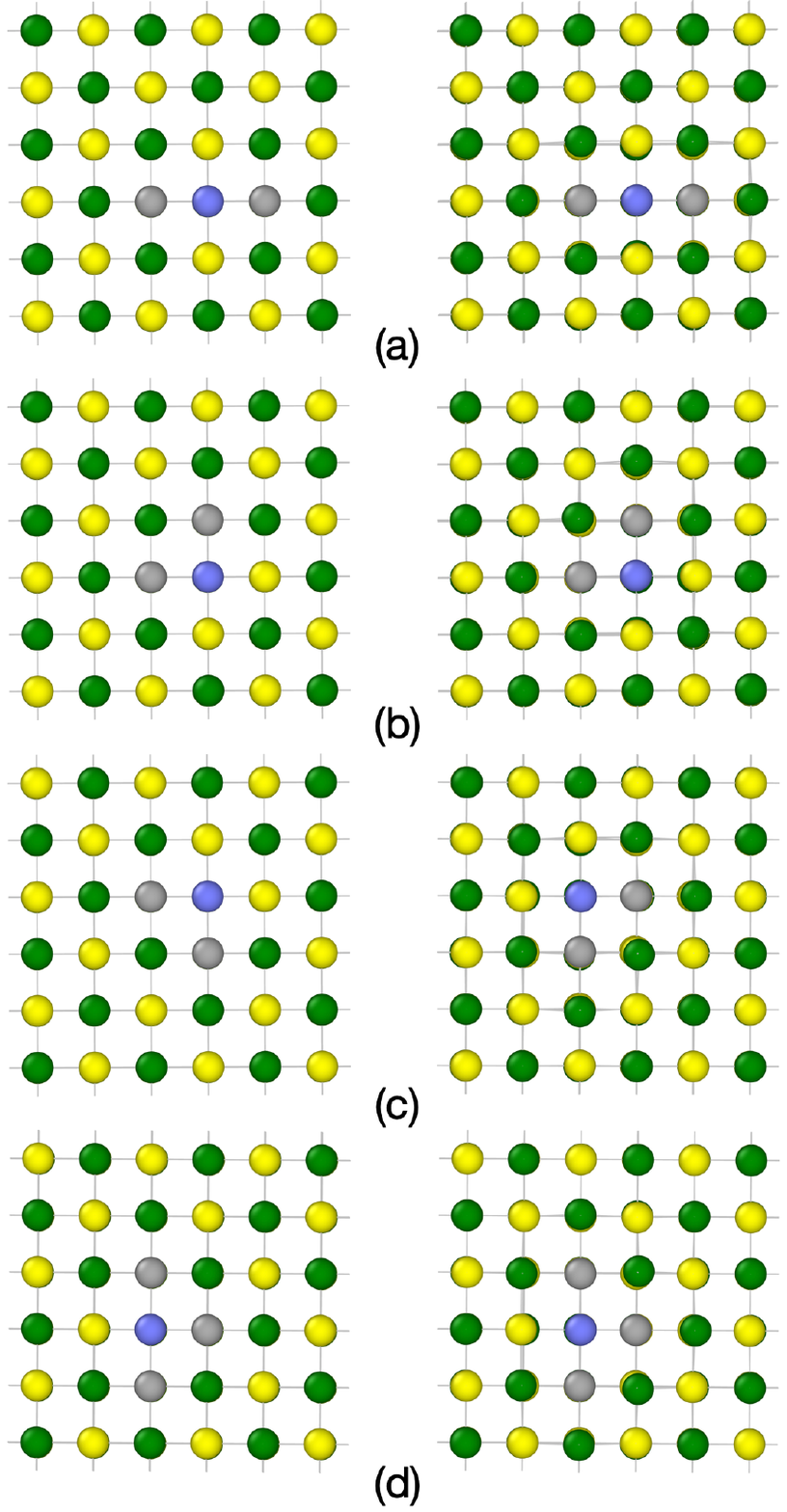}
\par\end{centering}
\caption{Initial (left column) and relaxed (right column) structures of point-defect
clusters undergoing structural reconstructions. (a) $V_{\alpha}V_{\beta}^{2}(L)$,
(b) $V_{\alpha}V_{\beta}^{2}(T)$ (c) $V_{\alpha}V_{\beta}^{3}(OP)$,
and (d) $V_{\alpha}V_{\beta}^{3}(IP)$ in TaC and HfC. The structures
are vie wed along a $\left\langle 100\right\rangle $ direction. The
metal atoms, carbon atoms, metal vacancies and carbon vacancies are
shown in green, yellow, purple, and light gray, respectively.\label{fig:relaxed_structures-1}}
\end{figure}

\begin{figure}
\begin{centering}
\includegraphics[width=0.8\textwidth]{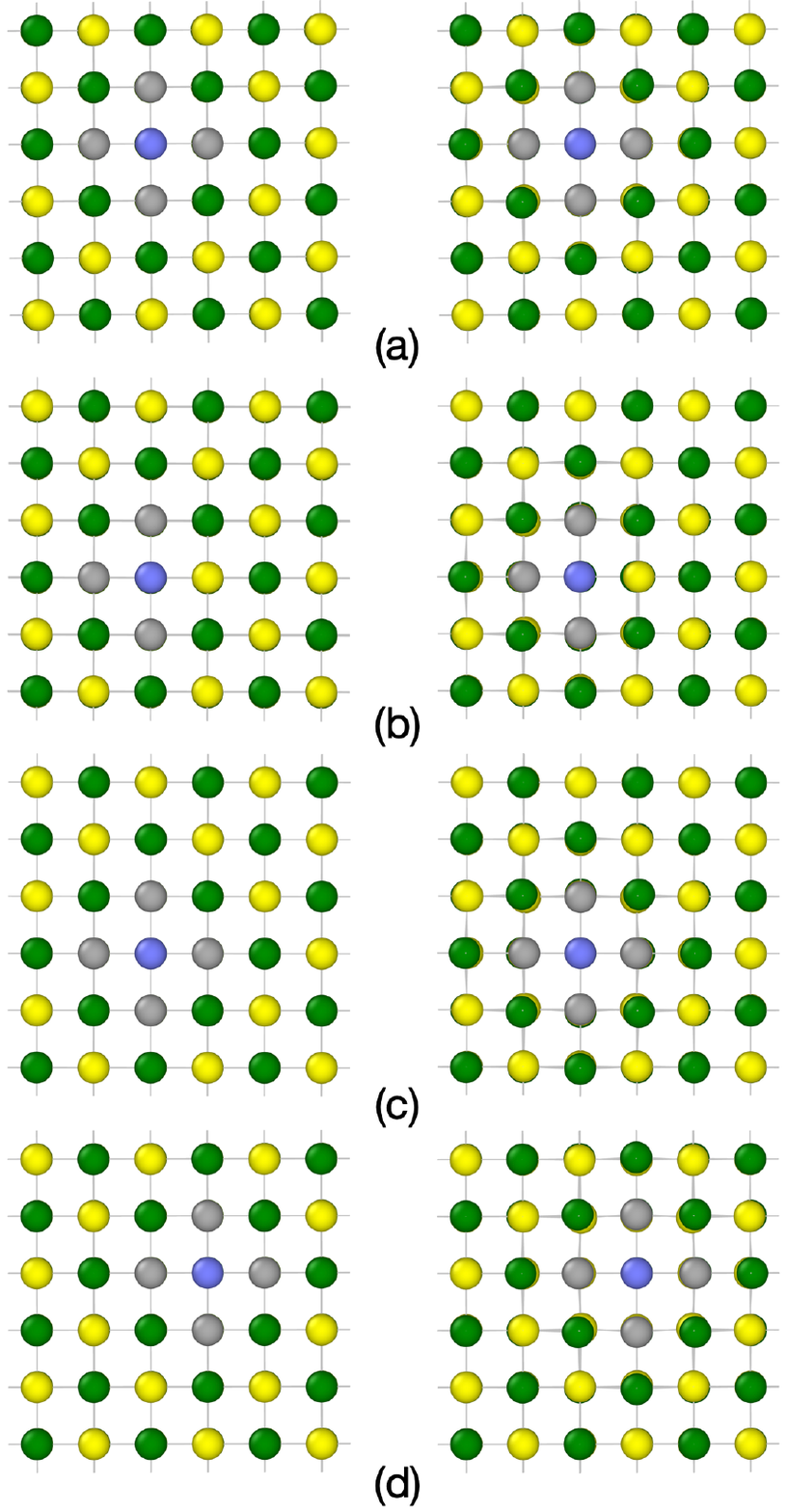}
\par\end{centering}
\caption{Initial (left column) and relaxed (right column) structures of point-defect
clusters undergoing structural reconstructions. (a) $V_{\alpha}V_{\beta}^{4}(IP)$,
(b) $V_{\alpha}V_{\beta}^{4}(OP)$, (c) $V_{\alpha}V_{\beta}^{5}$,
and (d) $V_{\alpha}V_{\beta}^{6}$ in TaC and HfC. The structures
are viewed along a $\left\langle 100\right\rangle $ direction. The
metal atoms, carbon atoms, metal vacancy and carbon vacancy are shown
in green, yellow, purple, and light gray, respectively.\label{fig:relaxed_structures-1}}
\end{figure}

\begin{figure}
\begin{centering}
\includegraphics[width=0.5\textwidth]{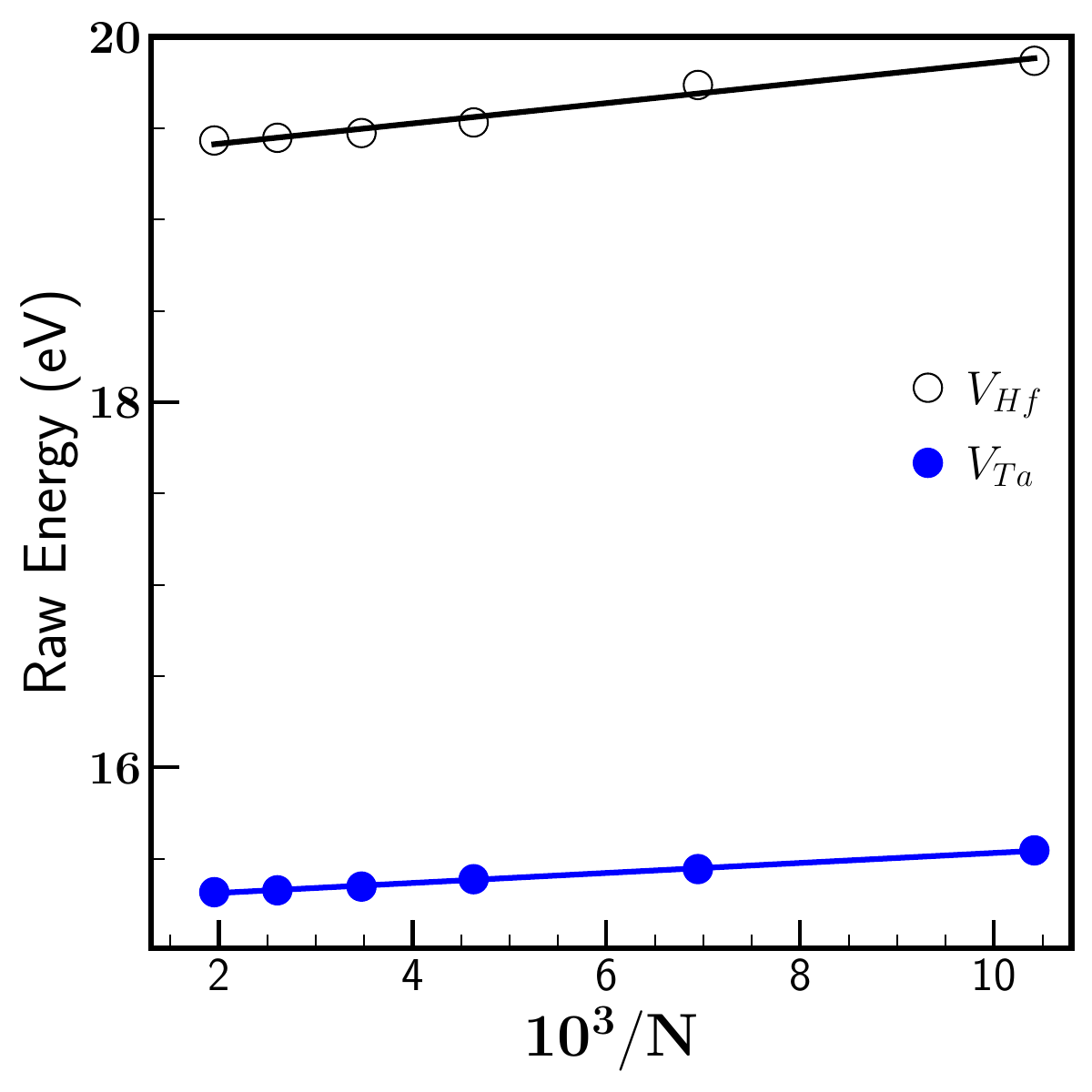}
\par\end{centering}
\caption{Raw energy of metal vacancy ($\epsilon_{V_{\alpha}}$) as the function
of reciprocal of the number of atoms prior to any defects for TaC
and HfC in rocksalt crystal structure.}
\end{figure}

\begin{figure}
\begin{centering}
\includegraphics[width=0.5\textwidth]{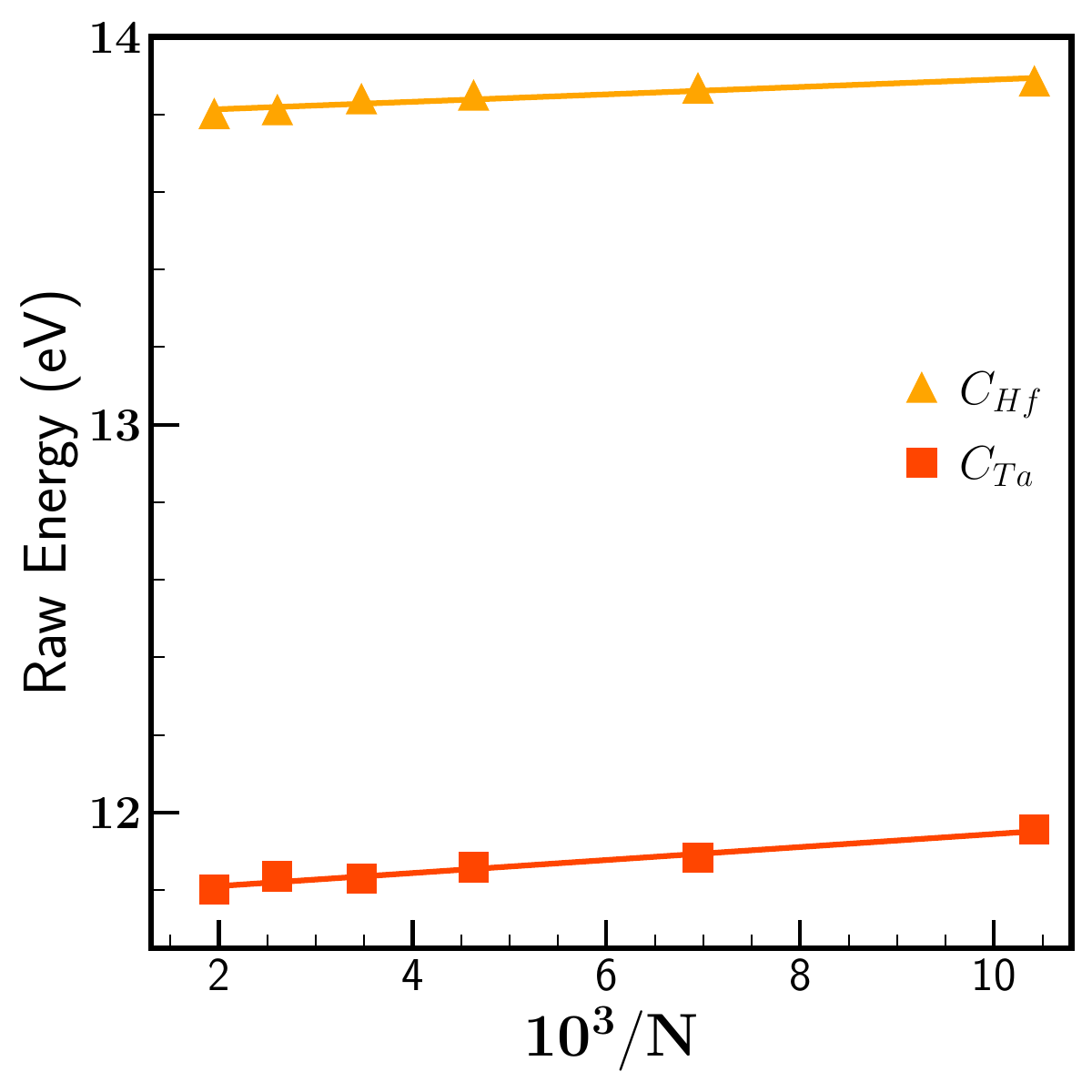}
\par\end{centering}
\caption{Raw energy of antisite on metal sublattice ($\epsilon_{B_{\alpha}}$)
as the function of reciprocal of the number of atoms prior to any
defects for TaC and HfC in rocksalt crystal structure.}
\end{figure}

\begin{figure}
\begin{centering}
\includegraphics[width=0.5\textwidth]{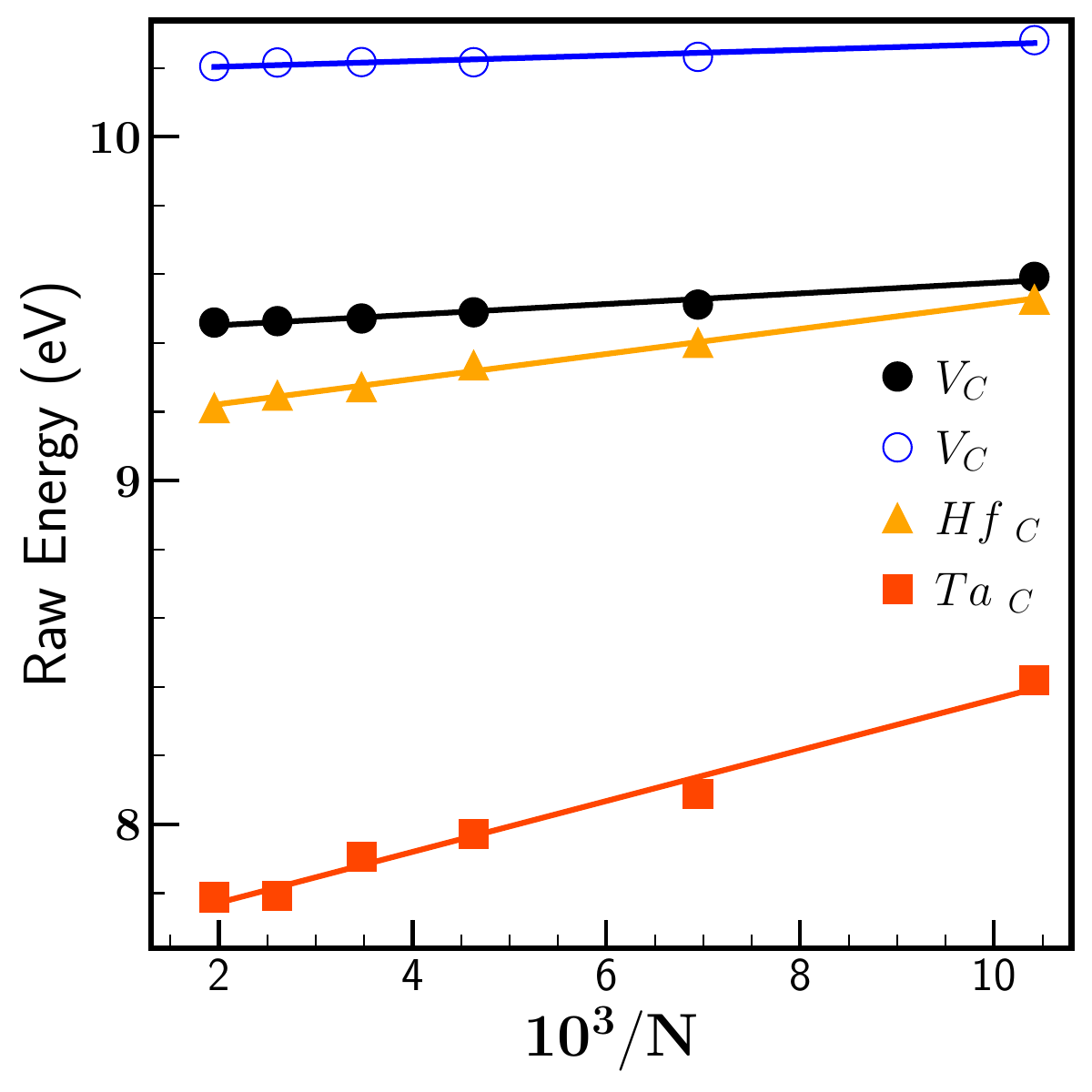}
\par\end{centering}
\caption{Raw energy of carbon vacancy ($\epsilon_{V_{\beta}}$) and antisite
on carbon sublattice ($\epsilon_{A_{\beta}}$) as a function of the
reciprocal of the number of atoms prior to any defects for TaC and
HfC in rocksalt crystal structure. Open and closed circles of carbon
vacancy ($V_{\beta}$) are for HfC and TaC respectively.}
\end{figure}

\begin{figure}
\begin{centering}
\includegraphics[width=0.5\textwidth]{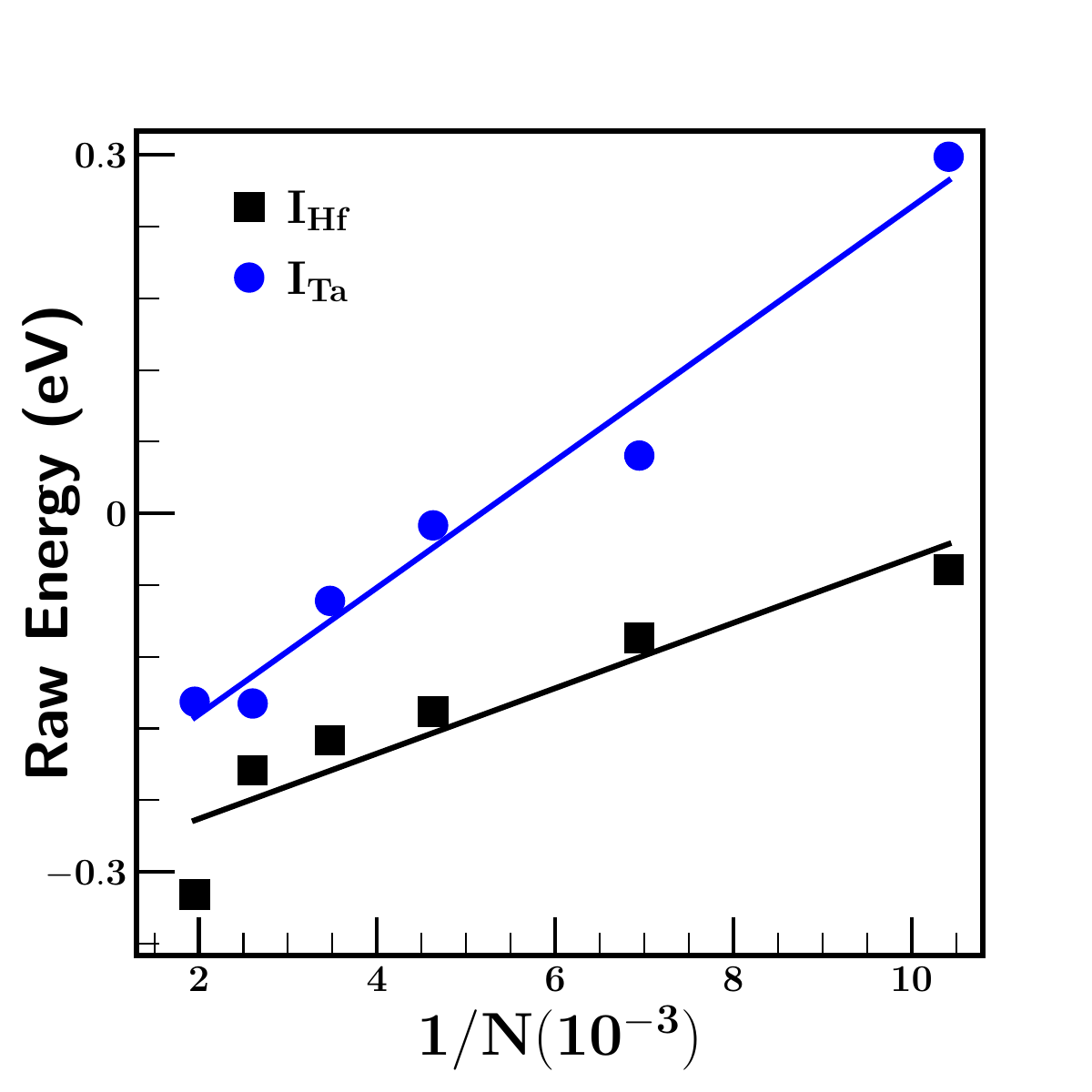}
\par\end{centering}
\caption{Raw energy of metal interstitial ($\epsilon_{I_{\alpha}}$) in tetrahedral
position as a function of the reciprocal of the number of atoms prior
to any defects for TaC and HfC in rocksalt crystal structure.}
\end{figure}

\begin{figure}
\begin{centering}
\includegraphics[width=0.5\textwidth]{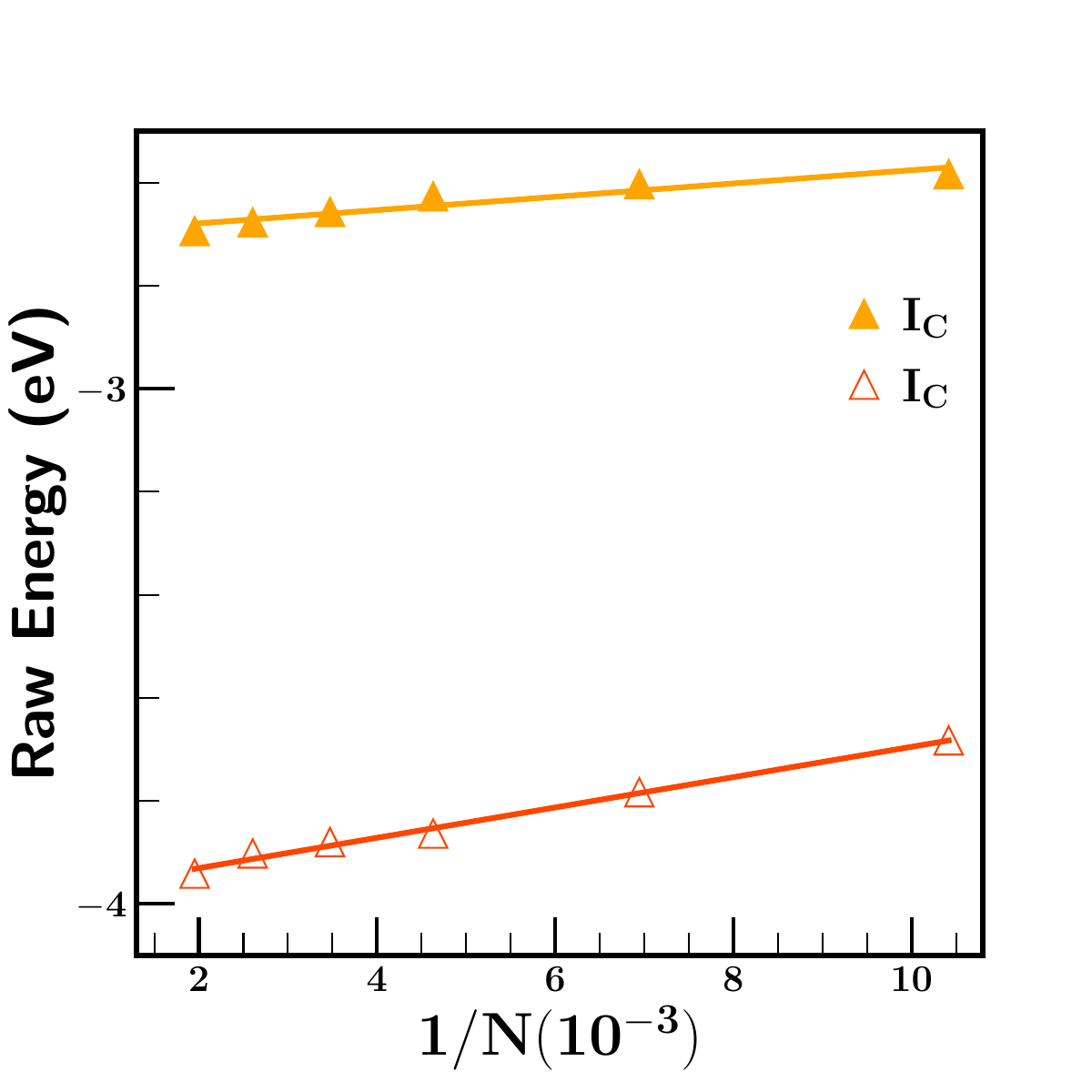}
\par\end{centering}
\caption{Raw energy of carbon interstitial ($\epsilon_{I_{\beta}}$) in tetrahedral
position as a function of the reciprocal of the number of atoms prior
to any defects for TaC and HfC in rocksalt crystal structure. Open
and closed triangles are for HfC and TaC respectively.}
\end{figure}

\begin{figure}
\begin{centering}
\includegraphics[width=0.5\textwidth]{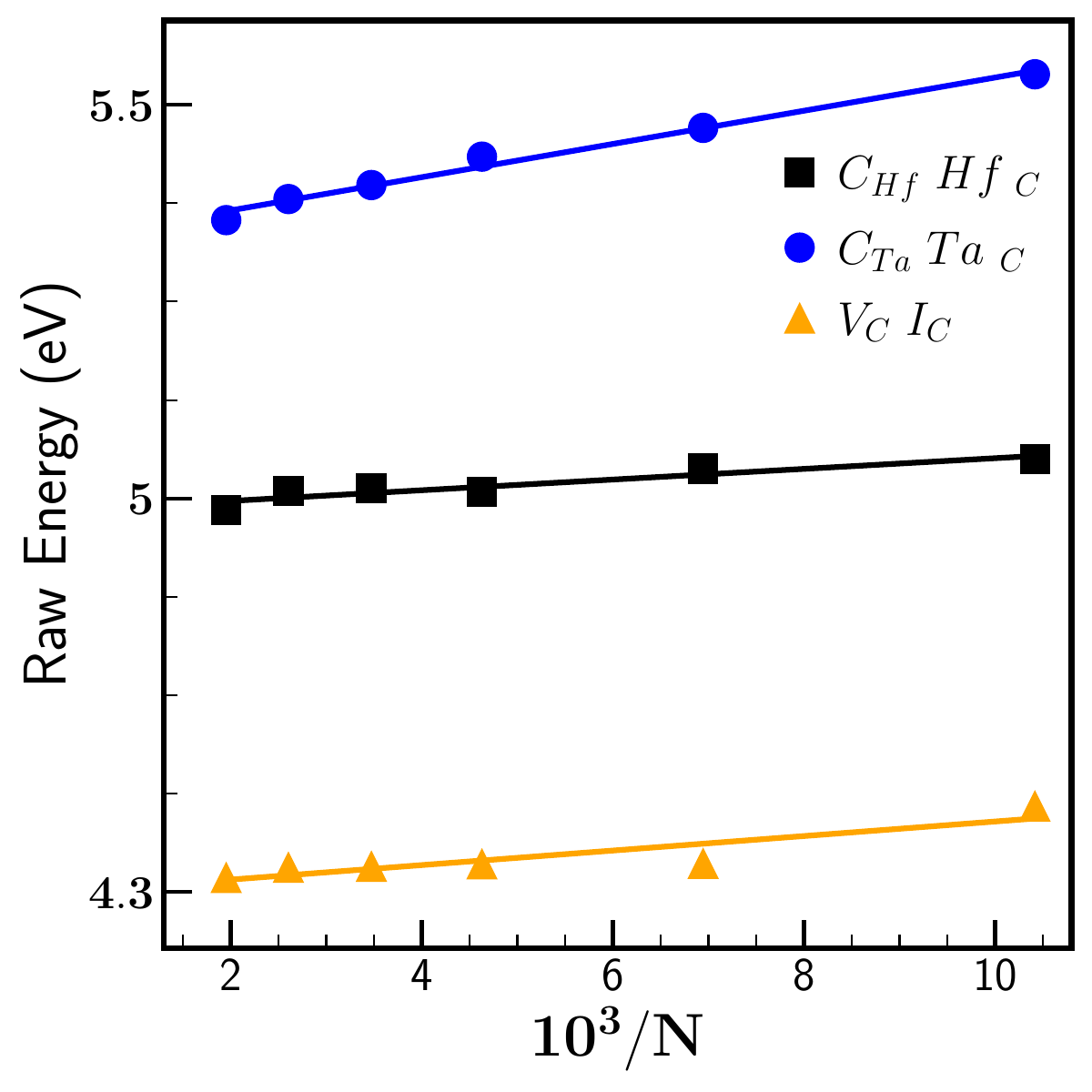}
\par\end{centering}
\caption{Raw energy of Frenkel pair ($\epsilon_{V_{\beta}I_{\beta}}$) in HfC
and antisite pair ($\epsilon_{B_{\alpha}A_{\beta}}$) in HfC and TaC
as a function of the reciprocal of the number of atoms prior to any
defects in rocksalt crystal structure.}
\end{figure}

\begin{figure}
\begin{centering}
\includegraphics[width=0.5\textwidth]{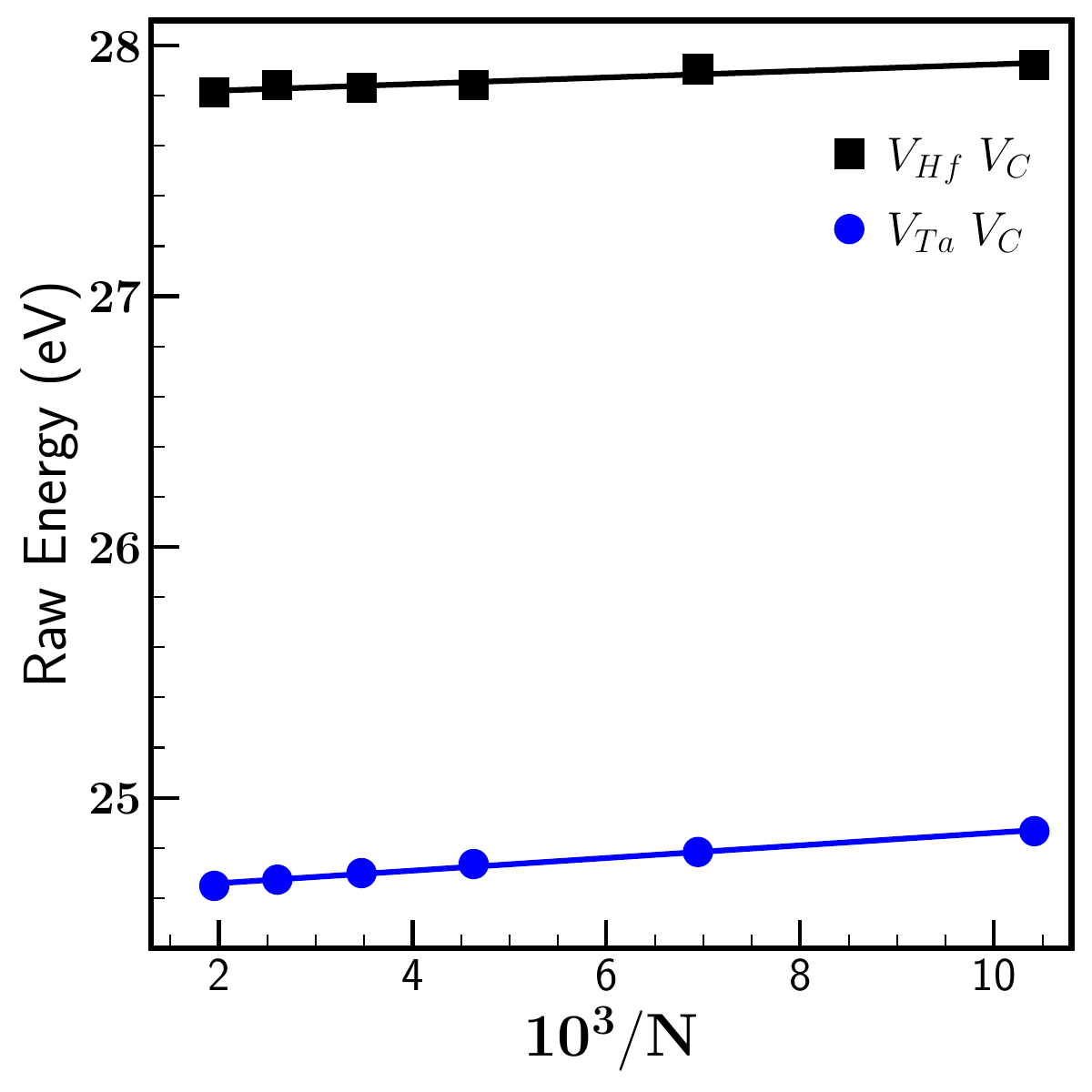}
\par\end{centering}
\caption{Raw energy of divacancy ($\epsilon_{V_{\alpha}V_{\beta}}$) as a function
of the reciprocal of the number of atoms prior to any defects for
TaC and HfC in rocksalt crystal structure.}
\end{figure}

\begin{figure}
\begin{centering}
\includegraphics[width=0.5\textwidth]{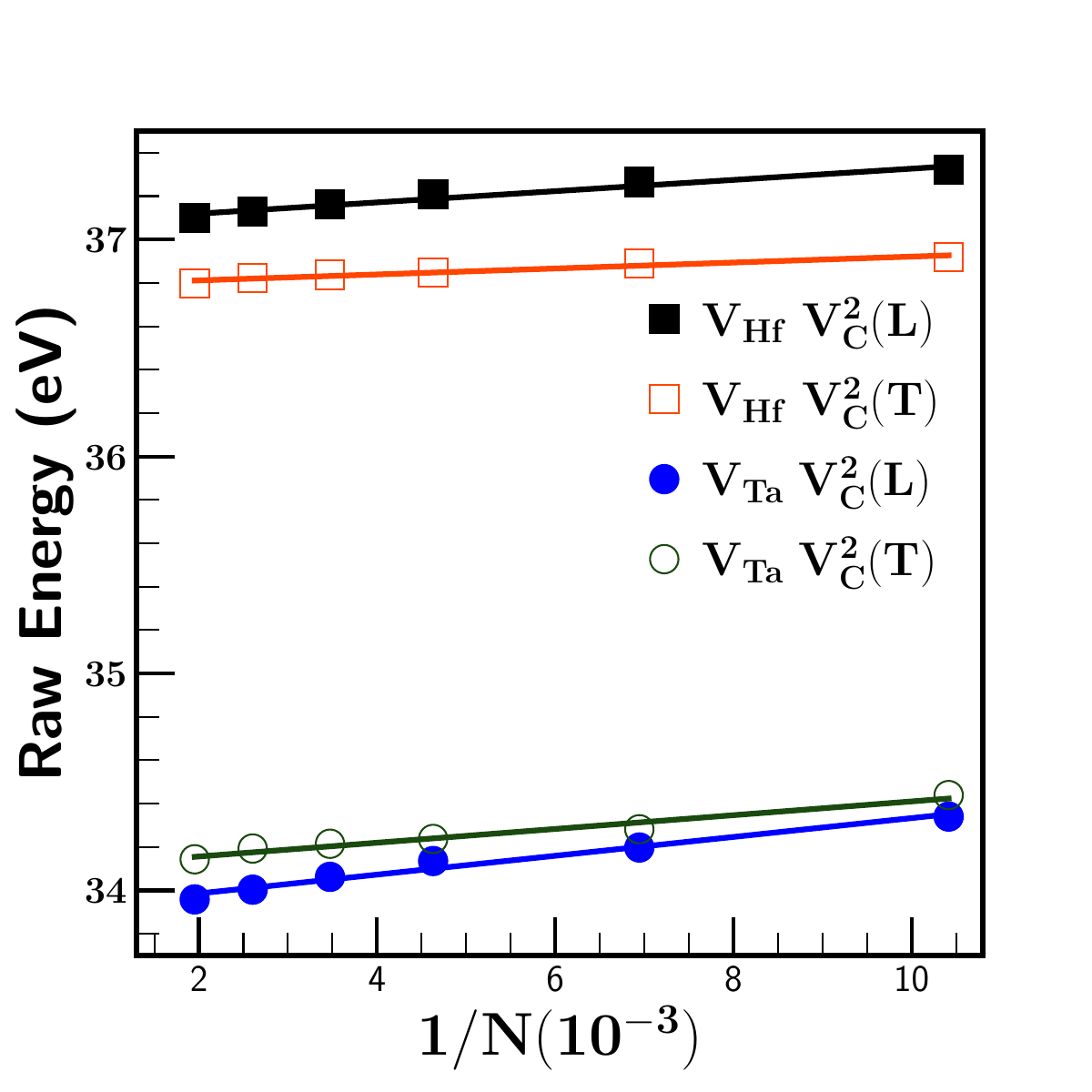}
\par\end{centering}
\caption{Raw energy of three vacancies cluster ($\epsilon_{V_{\alpha}V_{\beta}^{2}}$)
in Linear (L) and Triangular (T) configurations as a function of the
reciprocal of the number of atoms prior to any defects for TaC and
HfC in rocksalt crystal structure.}
\end{figure}

\begin{figure}
\begin{centering}
\includegraphics[width=0.5\textwidth]{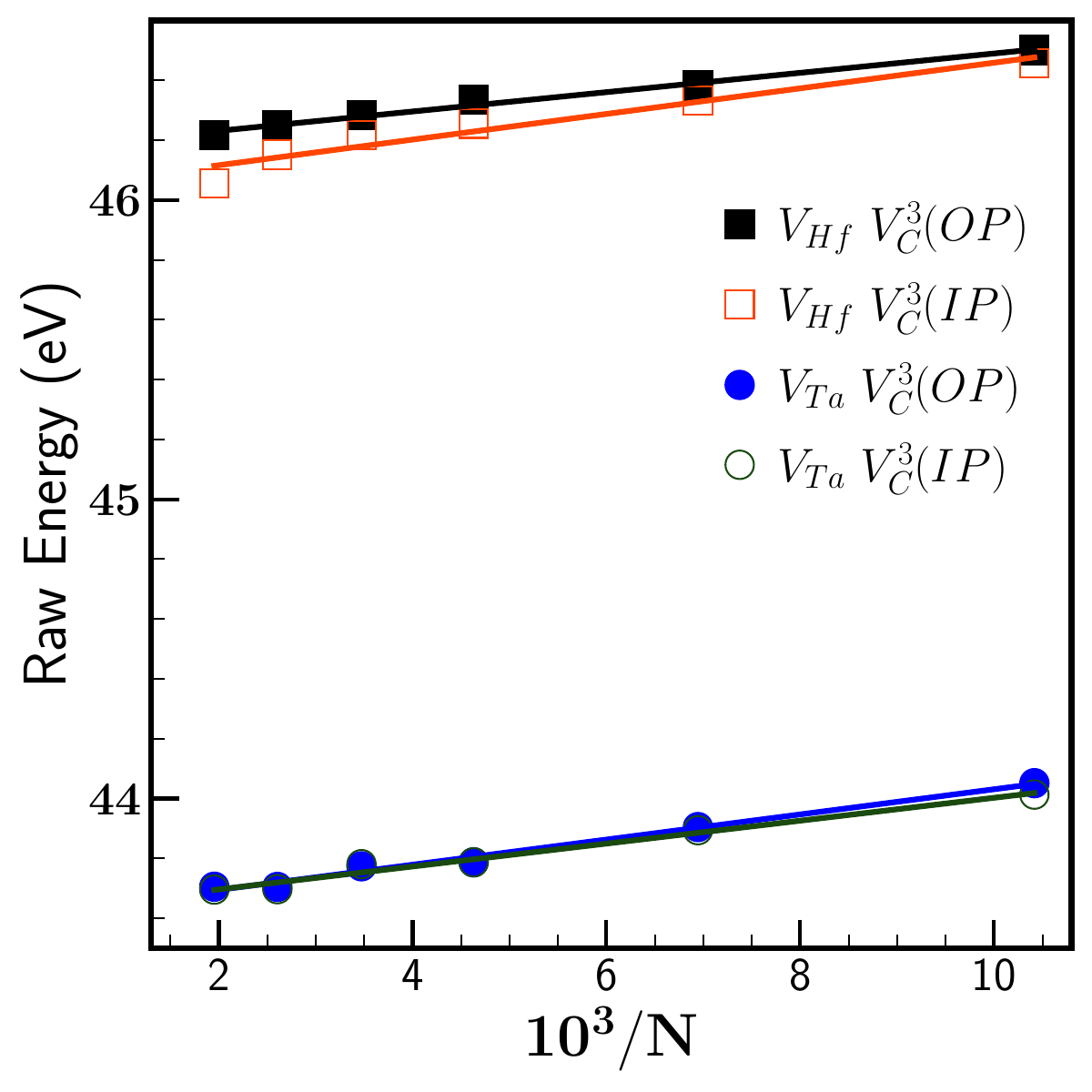}
\par\end{centering}
\caption{Raw energy of four vacancies cluster ($\epsilon_{V_{\alpha}V_{\beta}^{3}}$)
with three carbon vacancies are off-plane (OP) and in-plane (IP) configurations,
as a function of the reciprocal of the number of atoms prior to any
defects for TaC and HfC in rocksalt crystal structure.}
\end{figure}

\begin{figure}
\begin{centering}
\includegraphics[width=0.5\textwidth]{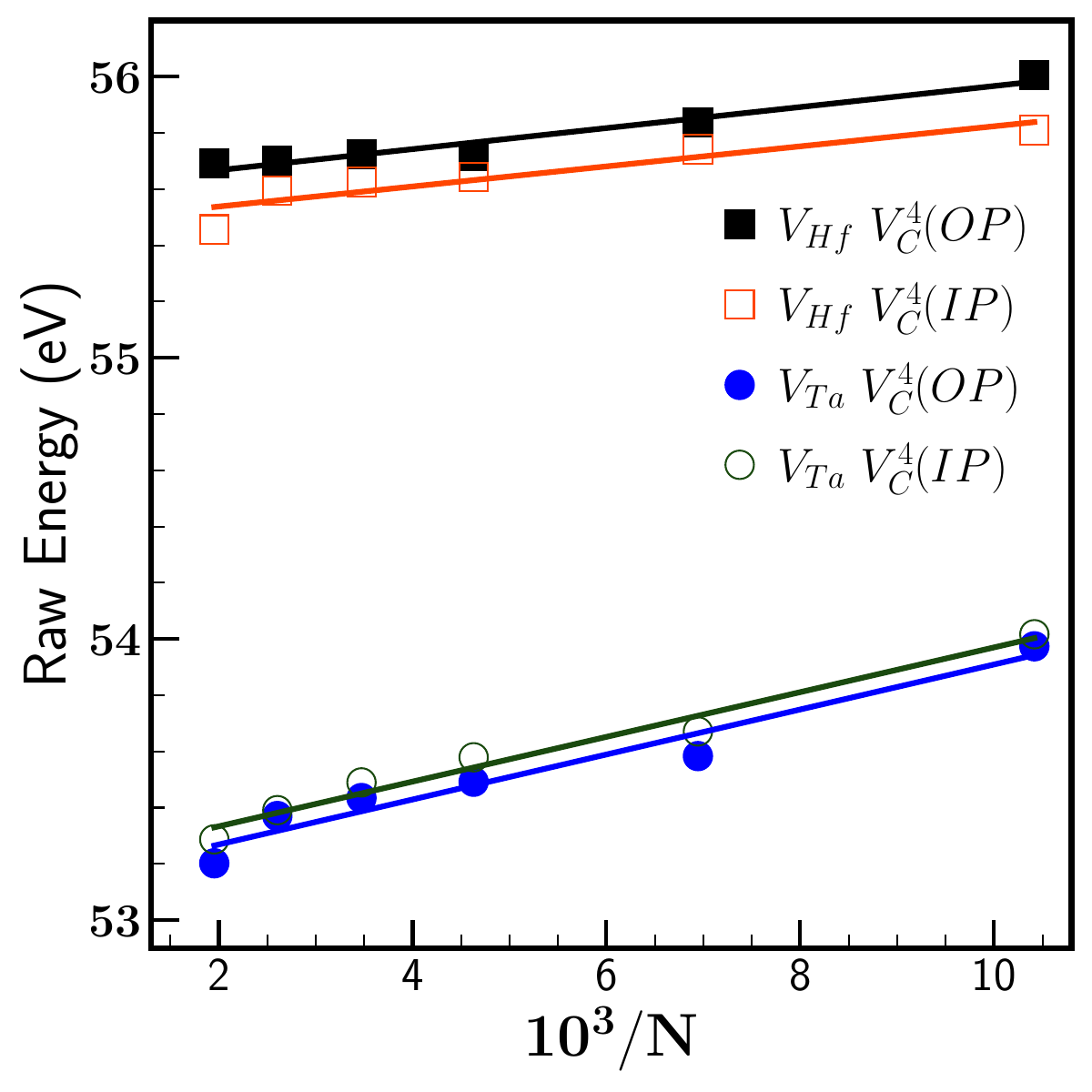}
\par\end{centering}
\caption{Raw energy of five vacancies cluster ($\epsilon_{V_{\alpha}V_{\beta}^{4}}$)
with four carbon vacancies are off-plane (OP) and in-plane (IP) configurations,
as a function of the reciprocal of the number of atoms prior to any
defects for TaC and HfC in rocksalt crystal structure.}
\end{figure}

\begin{figure}
\begin{centering}
\includegraphics[width=0.5\textwidth]{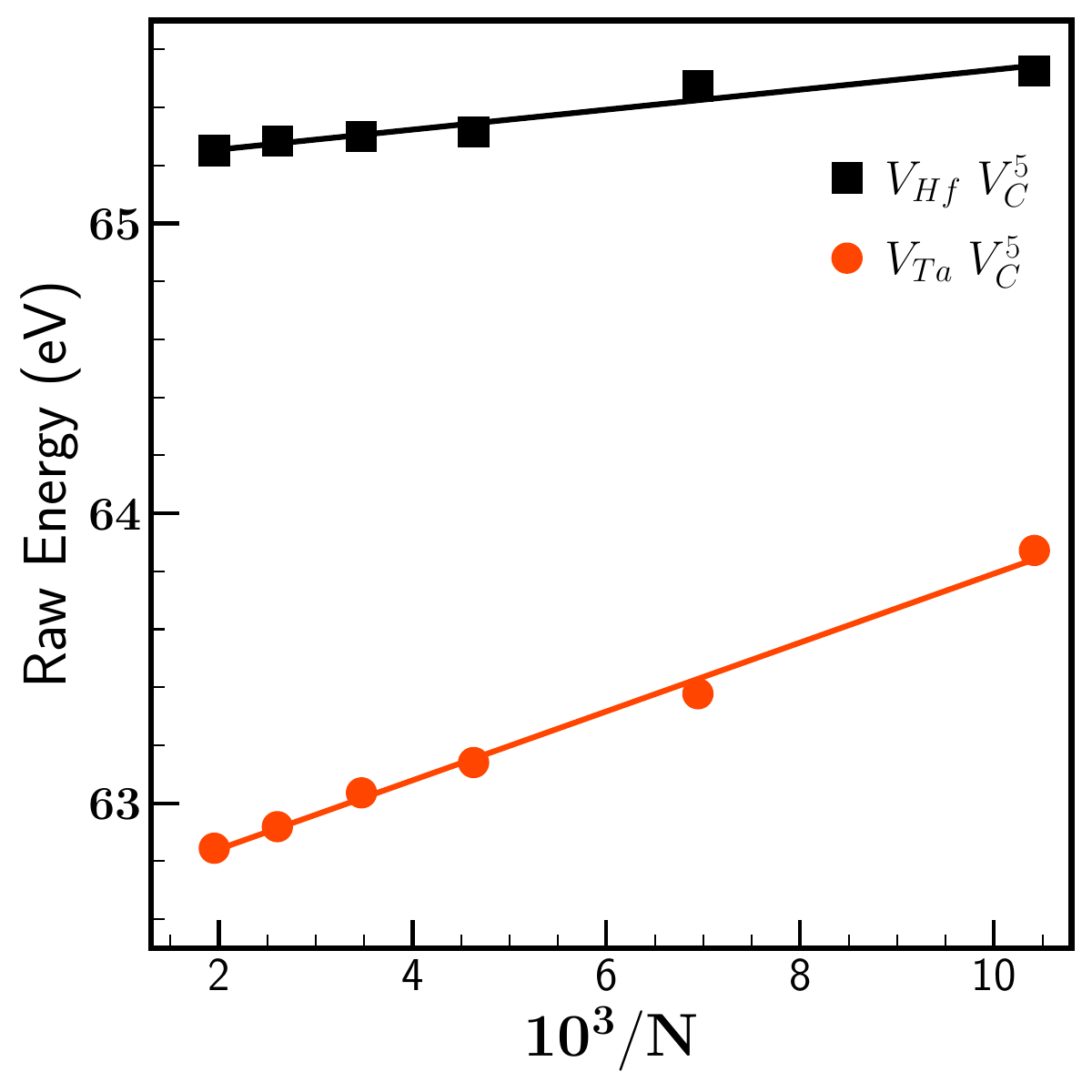}
\par\end{centering}
\caption{Raw energy of six vacancies cluster ($\epsilon_{V_{\alpha}V_{\beta}^{5}}$)
as a function of the reciprocal of the number of atoms prior to any
defects for TaC and HfC in rocksalt crystal structure.}
\end{figure}

\begin{figure}
\begin{centering}
\includegraphics[width=0.5\textwidth]{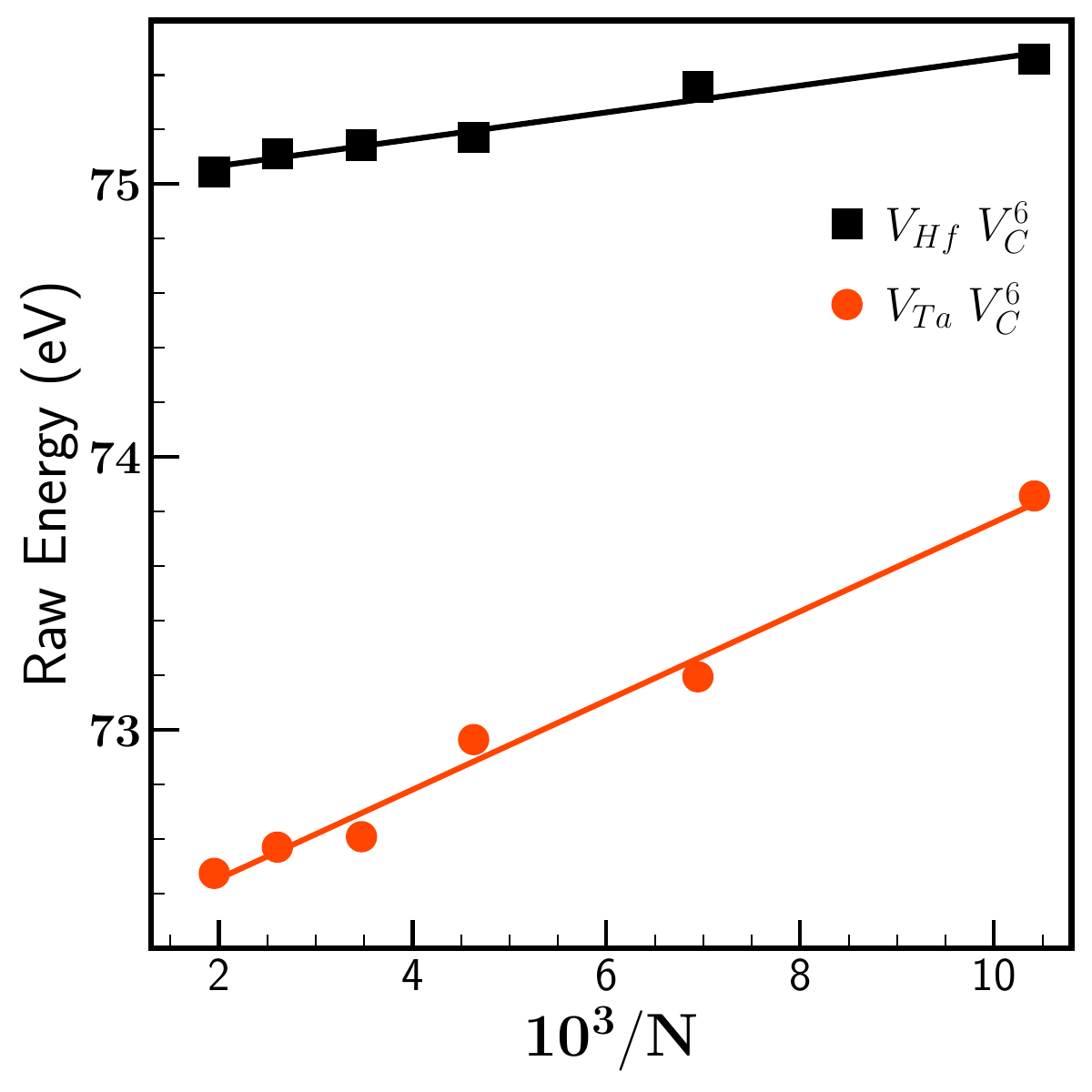}
\par\end{centering}
\caption{Raw energy of seven vacancies cluster ($\epsilon_{V_{\alpha}V_{\beta}^{6}}$)
as a function of the reciprocal of the number of atoms prior to any
defects for TaC and HfC in rocksalt crystal structure.}
\end{figure}

\begin{figure}
\begin{centering}
\includegraphics[width=0.5\textwidth]{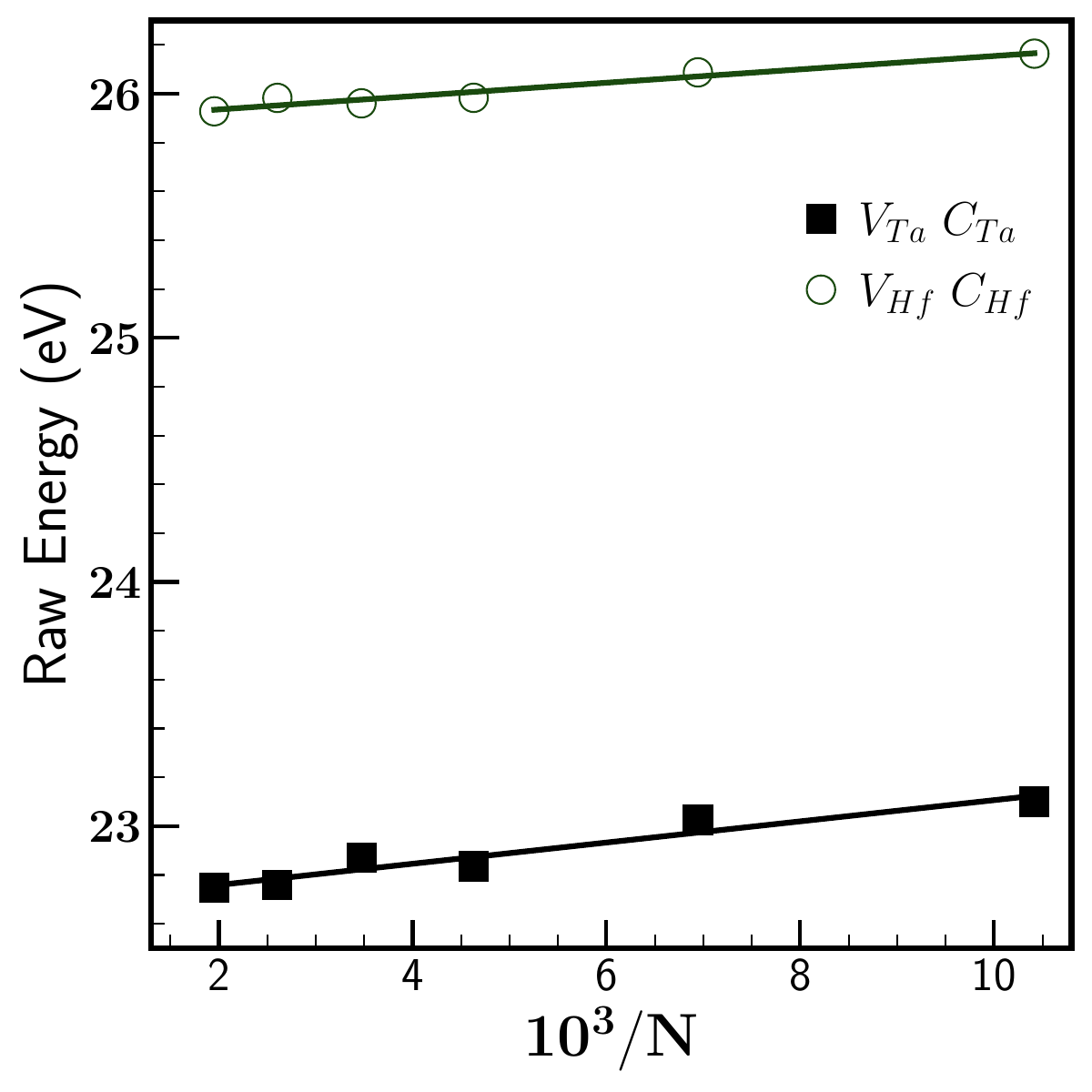}
\par\end{centering}
\caption{Raw energy of vacancy antisite pair on metal sublattice ($\epsilon_{V_{\alpha}B_{\alpha}}$)
as a function of the reciprocal of the number of atoms prior to any
defects for TaC and HfC in rocksalt crystal structure.}
\end{figure}

\begin{figure}
\begin{centering}
\includegraphics[width=0.5\textwidth]{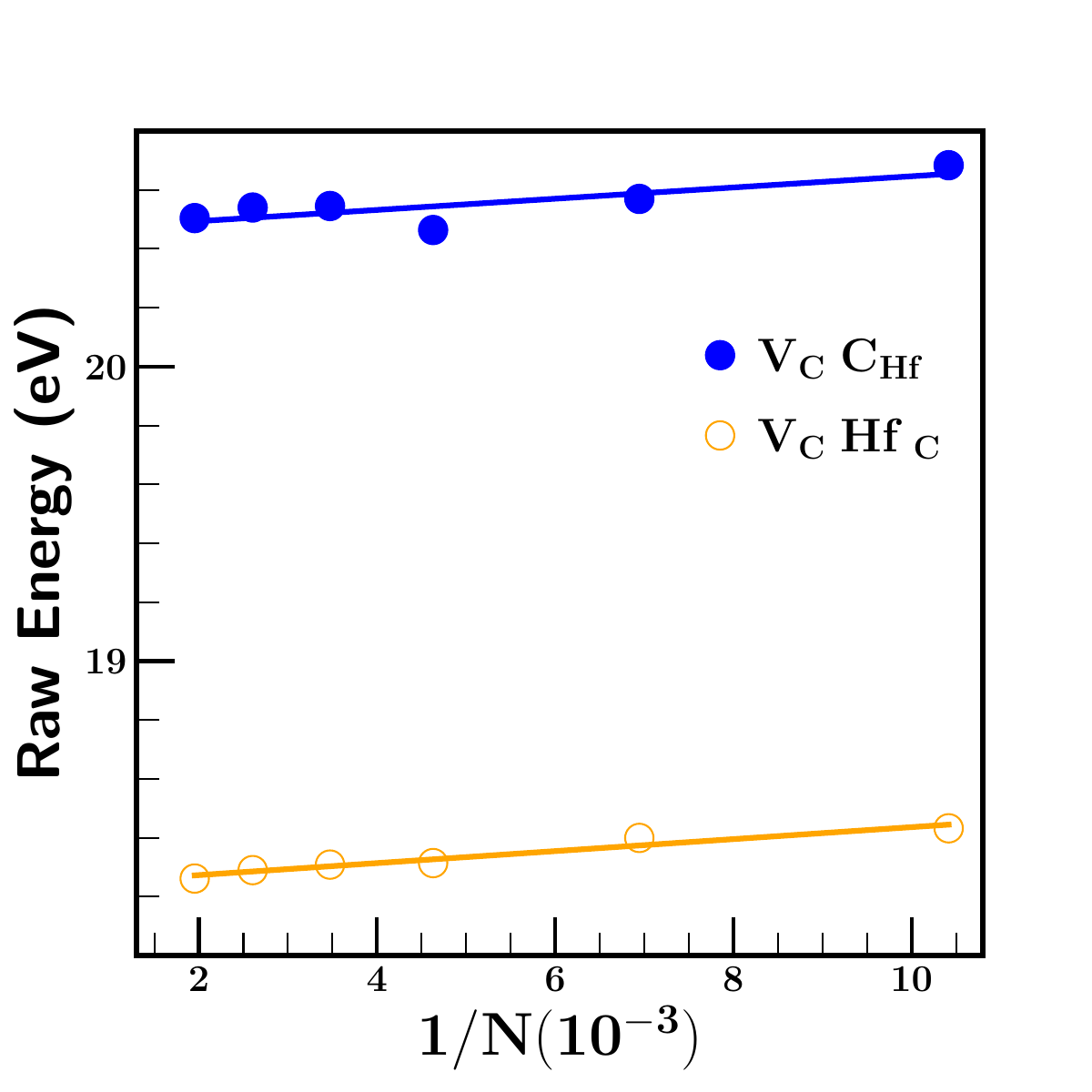}
\par\end{centering}
\caption{Raw energy of vacancy antisite pair on carbon sublattice ($\epsilon_{V_{\beta}A_{\beta}}$)
and carbon vacancy and antisite pair on metal sublattice ($\epsilon_{V_{\beta}B_{\alpha}}$)
in HfC as a function of the reciprocal of the number of atoms prior
to any defects in rocksalt crystal structure.}
\end{figure}

\begin{figure}
\begin{centering}
\includegraphics[width=0.5\textwidth]{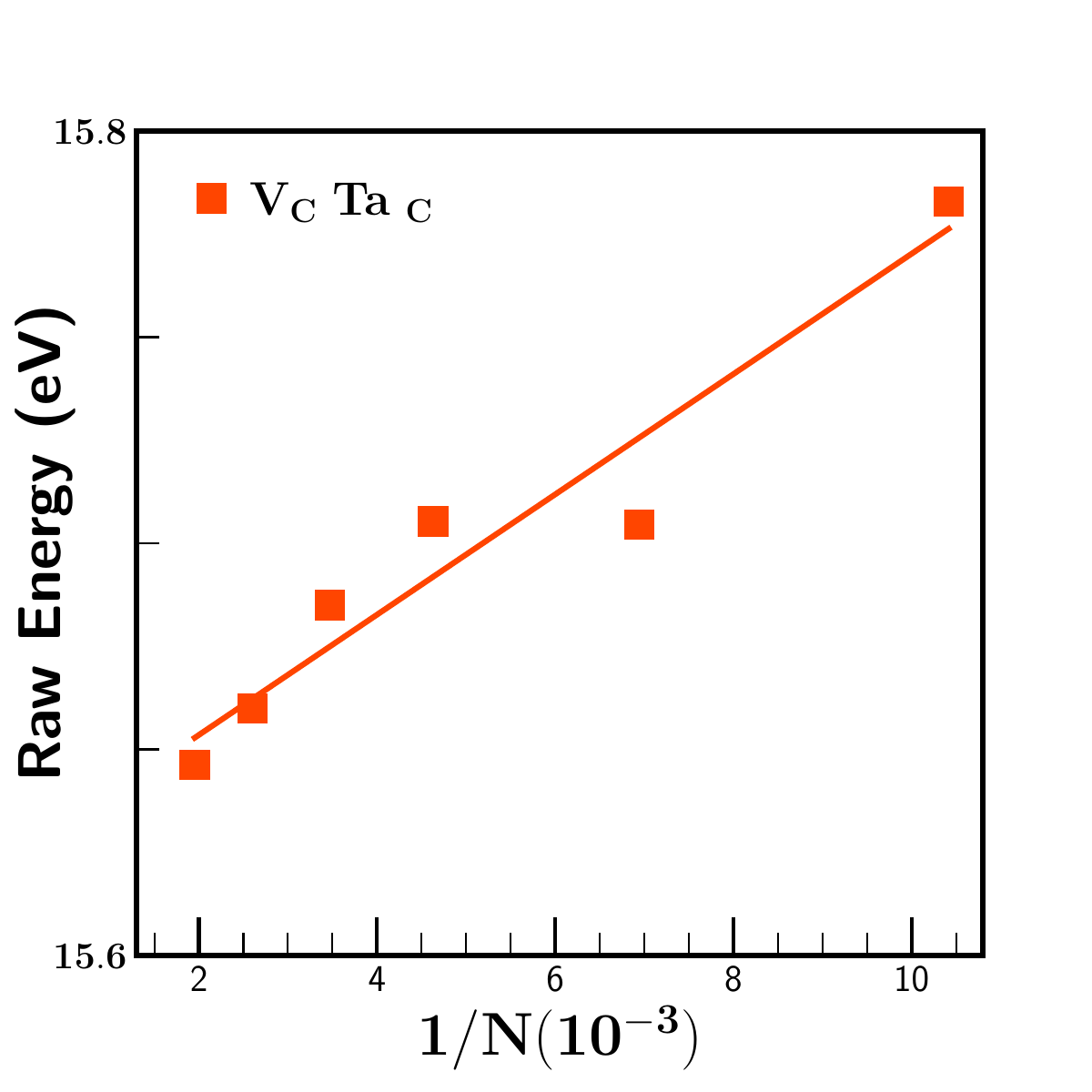}
\par\end{centering}
\caption{Raw energy of vacancy antisite pair on carbon sublattice ($\epsilon_{V_{\beta}A_{\beta}}$)
in TaC as a function of the reciprocal of the number of atoms prior
to any defects in rocksalt crystal structure.}
\end{figure}

\begin{figure}
(a)\includegraphics[width=0.47\textwidth]{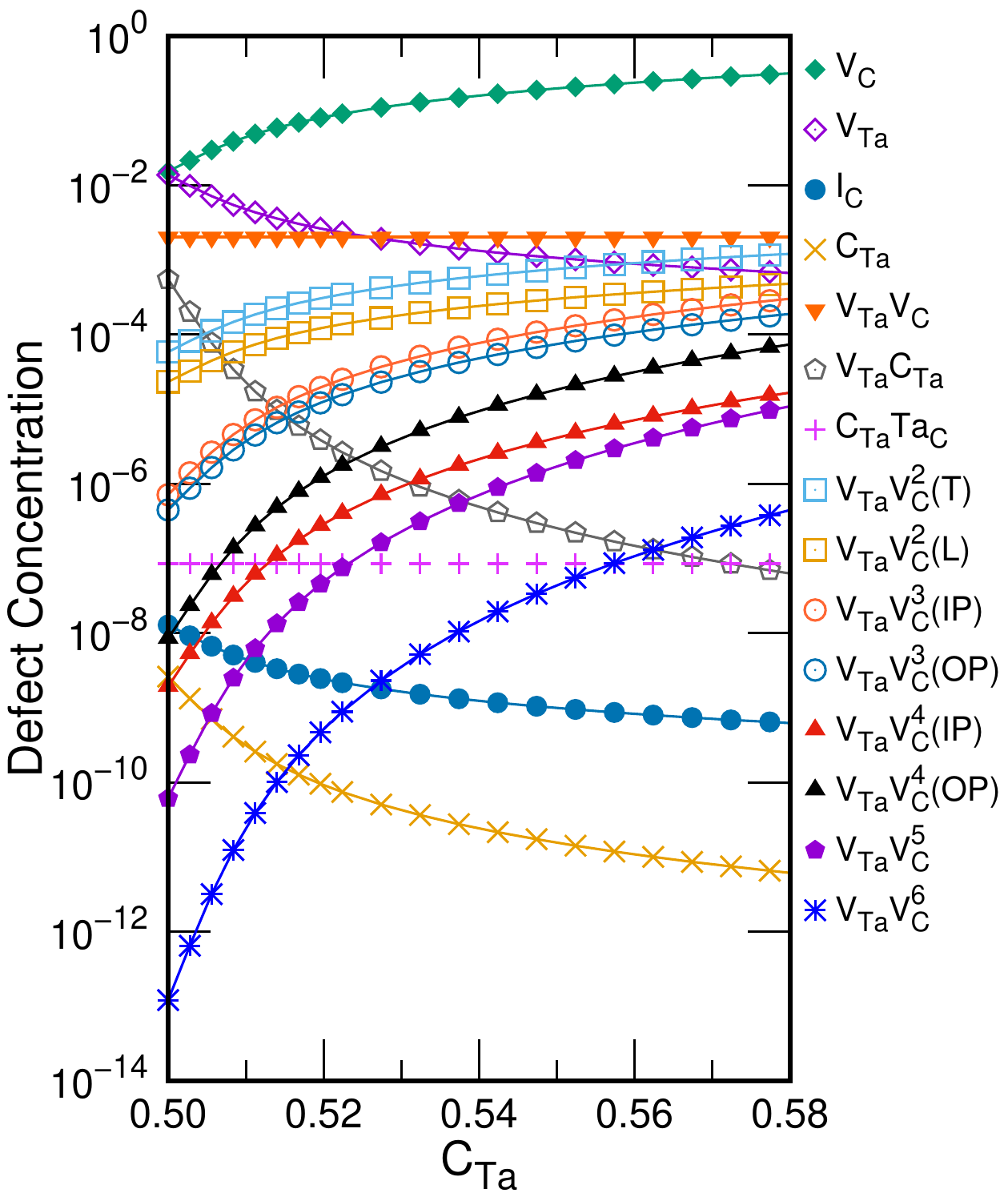}(b)\includegraphics[width=0.47\textwidth]{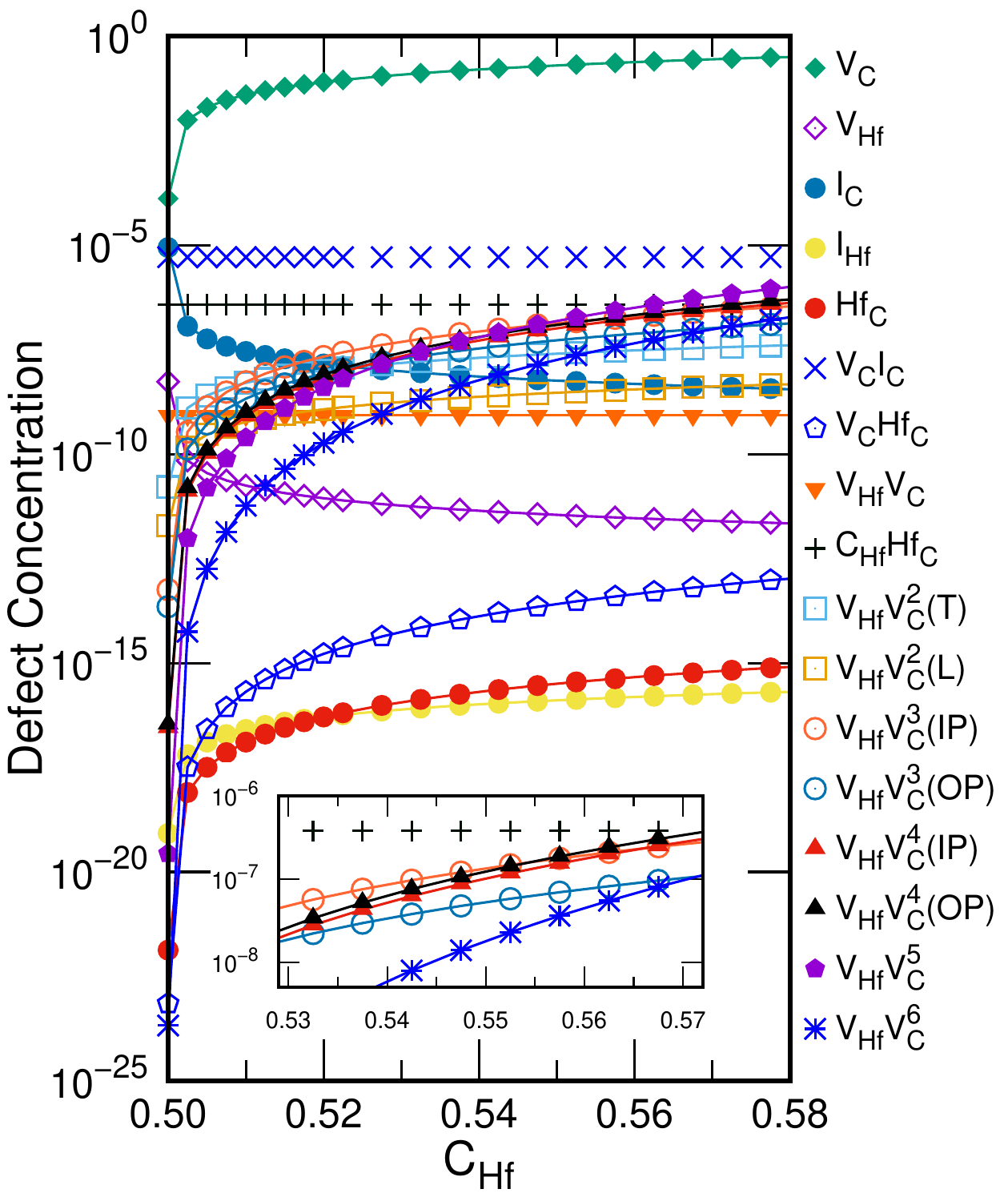}

\caption{Composition dependence of point defect concentrations in (a) TaC and
(b) HfC at 3500 K.\label{fig:Composition-dependence-1}}
\end{figure}

\begin{table}[!p]
\noindent \caption{The k-points used for raw energy calculations of various defects within
the supercells of HfC and TaC obtained through the repetition of the
relaxed conventional 8-atom unit cell of the B1 crystal structure.}
\bigskip{}
\begin{tabular}{|c|c|c|}
\hline 
Supercell Size (atoms) & K-points in TaC & K-points in HfC\tabularnewline
\hline 
$2\times2\times3$ (96) & $11\times11\times7$ & $2\times2\times1$\tabularnewline
\hline 
$3\times3\times2$ (144) & $7\times7\times11$ & $1\times1\times2$\tabularnewline
\hline 
$3\times3\times3$ (216) & $7\times7\times7$ & $3\times3\times3$\tabularnewline
\hline 
$3\times3\times4$ (288) & $6\times6\times4$ & $2\times2\times1$\tabularnewline
\hline 
$4\times4\times3$ (384) & $4\times4\times6$ & $1\times1\times2$\tabularnewline
\hline 
$4\times4\times4$ (512) & $5\times5\times5$ & $1\times1\times1$\tabularnewline
\hline 
\end{tabular}\label{table:kpoint}
\end{table}

\end{document}